\documentclass[twocolumn,numberedappendix,iop]{openjournal}
\usepackage{graphicx,amsmath,amssymb,amstext}
\usepackage{amsbsy,amsfonts,amsthm,color}
\usepackage[colorlinks,linkcolor=blue,citecolor=blue,urlcolor=blue ]{hyperref}
\usepackage[utf8]{inputenc}
\usepackage{float}
\usepackage{xcolor}
\usepackage{ulem}
\usepackage[T1]{fontenc}
\usepackage[title]{appendix}

\begin{document}

\title[BOSS Shear Ratio]{Cosmology with shear ratios: a joint study of weak lensing and spectroscopic redshift datasets \vspace{-4em}}

\author{Ni Emas$^{1,*}$}
\author{Chris Blake,$^{1}$}
\author{Rossana Ruggeri,$^{2,1}$}
\author{Anna Porredon$^{3}$}

\thanks{$^*$E-mail: nemas@swin.edu.au}
\affiliation{$^{1}$Centre for Astrophysics and Supercomputing, Swinburne University of Technology, P.O. Box 218, Hawthorn, VIC 3122, Australia\\
$^{2}$School of Mathematics and Physics, University of Queensland, Brisbane, QLD 4072, Australia\\
$^{3}$Ruhr University Bochum, Faculty of Physics and Astronomy, Astronomical Institute, German Centre for Cosmological Lensing, 44780 Bochum, Germany
}

\begin{abstract}
The ratio of the average tangential shear signal of different weak lensing source populations around the same lens galaxies, also known as a shear ratio, provides an important test of lensing systematics and a potential source of cosmological information.  In this paper we measure shear ratios in three current weak lensing surveys -- the Kilo-Degree Survey, Dark Energy Survey  and Subaru Hyper Suprime-Cam survey -- using overlapping data from the Baryon Oscillation Spectroscopic Survey.  We apply a Bayesian method to reduce bias in shear ratio measurement, and assess the degree to which shear ratio information improves the determination of important astrophysical parameters describing the source redshift distributions and intrinsic galaxy alignments, as well as cosmological parameters, in comparison with cosmic shear and full $3 \times 2$-pt correlations (cosmic shear, galaxy-galaxy lensing, and galaxy clustering).  We consider both Fisher matrix forecasts, as well as full likelihood analyses of the data.  We find that the addition of shear ratio information to cosmic shear allows the mean redshifts of the source samples and intrinsic alignment parameters to be determined significantly more accurately.  Although the additional constraining power enabled by the shear ratio is less than that obtained by introducing an accurate prior in the mean source redshift using photometric redshift calibration, the shear ratio allows for a useful cross-check.  The inclusion of shear ratio data consistently benefits the determination of cosmological parameters such as $S_8 = \sigma_8 \sqrt{\Omega_\mathrm{m}/0.3}$, for which we obtain improvements up to 34\%. However these improvements are less significant when shear ratio is combined with the full $3 \times 2$-pt correlations.  We conclude that shear ratio tests will remain a useful source of cosmological information and cross-checks for lensing systematics, whose application will be further enhanced by upcoming datasets such as the Dark Energy Spectroscopic Instrument.
  \\[1em]
  \textit{Keywords:} gravitational lensing: weak -- large-scale structure of Universe -- methods: statistical -- cosmological parameters
\end{abstract}

\maketitle

\section{Introduction}

We have entered an era of precision cosmology, in which multiple parameters that define the cosmological model may be accurately measured by combinations of cosmological probes.  Two parameters that have attracted significant interest in recent years are the Hubble constant $H_0$, which defines the local cosmic expansion rate, and a parameter that describes the amount and clustering of matter in the Universe, $S_8 = \sigma_8 \sqrt{\Omega_\mathrm{m}/0.3}$, where $\sigma_8$ is the amplitude of matter fluctuations on a scale of 8 $h^{-1}$Mpc, and $\Omega_\mathrm{m}$ is the matter density parameter.

The $S_8$ parameter has come into focus because it may be accurately determined by weak gravitational lensing \citep[e.g.,][]{Mandelbaum2017WeakLF}, in a complementary manner to early-Universe predictions.  Weak lensing measures a statistical signal by averaging over the correlated shape distortions of a large number of source galaxies in imaging surveys, as their light passes through the cosmic web and near foreground lens galaxies.  There is a statistical tension of 2-3$\sigma$ between $S_8$ measurements from some weak lensing surveys \citep{2018MNRAS.474.4894J, 2021A&A...646A.140H, 2023arXiv230400704M} and standard model fits to the Cosmic Microwave Background (CMB) \citep{2020A&A...641A...6P}, in which the CMB prediction of $S_8$ is higher than that obtained using model fits to the lensing datasets. There is also some evidence that other probes of the late universe such as spectroscopic galaxy clustering \citep{Tr_ster_2021}, redshift-space distortion (RSD) \citep{BENISTY2021100766}, and galaxy clusters \citep{Mantz_2014}, find lower $S_8$ results than the measurement from the CMB. This raises the question of the cause of this apparent tension: is it due to astrophysical systematic effects or new physics?

Weak lensing and galaxy survey datasets can be used to determine three different types of two-point (2-pt) correlations between the lenses and sources.  Cosmic shear describes the correlated distortion of the shapes of distant source galaxies due to weak lensing, galaxy clustering studies the correlations between lens galaxy positions, and galaxy-galaxy lensing (GGL) measures the cross-correlation between lens galaxy positions and source galaxy shapes, providing a link between galaxy clustering and cosmic shear.  This suite of observable statistics is known as the $3 \times 2$-pt correlations.  In order to use these measurements to extract cosmological parameters, we also need to model several significant systematic astrophysical effects that are not well-determined: the source redshift distributions, intrinsic alignments, shear calibration systematics, galaxy bias (how galaxies trace matter), and baryonic effects on the power spectrum at small scales \citep{2018ARA&A..56..393M}.

A complementary observational probe to the $3 \times 2$-pt correlation functions, which has been proposed to reduce some of these systematic effects, is the shear ratio.  This observable consists of a ratio of the tangential shear signal from GGL of two different weak lensing source samples around the same lens distribution. This ratio may partially cancel the effect of the mass or galaxy bias properties associated with the lens, potentially allowing smaller scales (where the signal is strongest) to be utilised, and maximizing the sensitivity to the geometrical information of the source survey, such as its redshift distribution.

Several previous studies have discussed and applied the shear ratio technique.  \cite{Jain&Taylor} proposed that the shear ratio could be used to constrain cosmological parameters, focusing on the equation of state of dark energy.  \cite{Taylor2007} quantified the constraints on cosmological parameters through this method, and further refined these suggestions by proposing to apply the ratio technique behind clusters using individual shear measurements rather than correlation functions. According to \cite{Taylor2007} and \cite{Bernstein&Jain2004}, the sensitivity of the shear ratio to cosmological parameters might not be strong, unless the redshift distribution and calibration of shear biases were accurately known. \cite{Kitching2008} considered the geometric shear ratio in the context of systematic parameters such as the source redshift distribution uncertainty due to photometric redshifts, shear calibration systematics, and intrinsic alignments (IA), confirming that astrophysical nuisance parameters greatly influenced the determination of cosmological parameters.  In this case, lensing ratios of galaxy-galaxy lensing measurements have been established as a probe to test redshift distributions and redshift-dependent multiplicative bias \citep{2016A&A...592L...6S} and the self-consistency of the shear-redshift calibration \citep{2021A&A...645A.105G}.  This method is known as the ``shear-ratio'' (SR) test.  If this probe is combined with other data such as $3 \times 2$-pt correlations, it can still improve measurements of cosmological parameters, as demonstrated by the analysis of \citet*{2022PhRvD.105h3529S} using data from the Dark Energy Survey.

In this work we extend the shear-ratio test method of \citet*{2022PhRvD.105h3529S} to the latest weak lensing and galaxy datasets, obtaining new constraints on cosmological and nuisance parameters.  By using spectroscopic redshift data for the galaxy lenses rather than photometric data as used by \citet*{2022PhRvD.105h3529S}, we avoid potential systematic uncertainties and nuisance parameters associated with the lenses, such as the imprint of inhomogeneities in photometric lens selection over the sky, or uncertainties in the lens redshift distributions.  We also use the Fisher matrix technique to supply, for the first time, detailed forecasts for the potential of shear ratio measurements to contribute additional information to shear-galaxy correlation functions.

The methodology of \citet*{2022PhRvD.105h3529S} uses a full model of GGL, including the corresponding integration over the power spectrum and the contribution from IA, source redshift distributions and lens magnification, allowing these astrophysical nuisance parameters to be fully modelled and constrained.  We use data from SDSS-III's Baryon Oscillation Spectroscopic Survey \citep[BOSS,][]{BOSS_1} as the lenses, along with three overlapping weak lensing surveys: the Kilo-Degree Survey (KiDS), adopting the KiDS-1000 shape catalogues \citep{2021A&A...645A.105G}, the Dark Energy Survey (DES) Year 3 \citep*{2021MNRAS.504.4312G} and the Subaru Hyper Suprime-Cam (HSC) Year 1 \citep{2018PASJ...70S..25M}. All these datasets are publicly available, and their combination allows us to perform statistical analysis of $3 \times 2$-pt correlation functions and shear ratios.

In the near future, the Dark Energy Spectroscopic Instrument (DESI) \citep{2016arXiv161100036D} will provide a new galaxy redshift dataset overlapping with the three weak lensing surveys, creating new opportunities for $3 \times 2$-pt analysis and shear ratio tests.  The next generation of weak lensing surveys will further enhance these possibilities, such as the {\it Euclid} satellite \footnote{\url{https://sci.esa.int/web/euclid}}, the Rubin Observatory’s Legacy Survey of Space and Time (LSST)\footnote{\url{https://www.lsst.org/}}, and the Roman Space Telescope\footnote{\url{https://roman.gsfc.nasa.gov/}}. The advent of this suite of observatories will provide highly-precise diagnostics for the parameter tensions in cosmology.

The structure of our paper is as follows.  Section \ref{sec:data} describes the galaxy and weak lensing survey data used in our analysis: BOSS, KIDS-1000, DES-Y3 and HSC-Y1.  In Section \ref{sec:measurements} we discuss our measurements of the average tangential shear and shear ratio, including the Bayesian method we implement to robustly measure shear ratios.  Section \ref{sec:shearmodel} describes the averaged tangential shear model and parameterisation used in this work, including both cosmological and nuisance parameters. We present Fisher matrix forecasts in Section \ref{sec:fisher}, and Bayesian fitting of cosmological and astrophysical parameters in Section \ref{sec:mcmc}.  We conclude our work in Section \ref{sec:conc}.

\section{Data}
\label{sec:data}

\subsection{BOSS}
\begin{figure}
  \includegraphics[width=\linewidth]{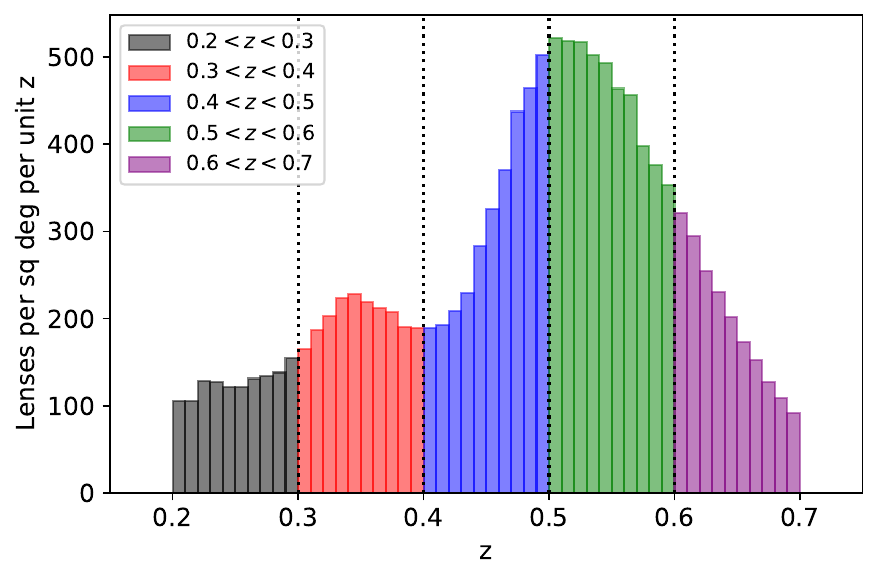}
  \caption{The redshift distribution of BOSS galaxies used in our analysis, plotted as the number of galaxies per square degree per unit redshift. We split the dataset into five lens bins in the range $0.2 < z < 0.7$ with width $\Delta z = 0.1$.}
  \label{fig:nz_lens}
\end{figure}

\begin{figure*}
  \includegraphics[width=\linewidth]{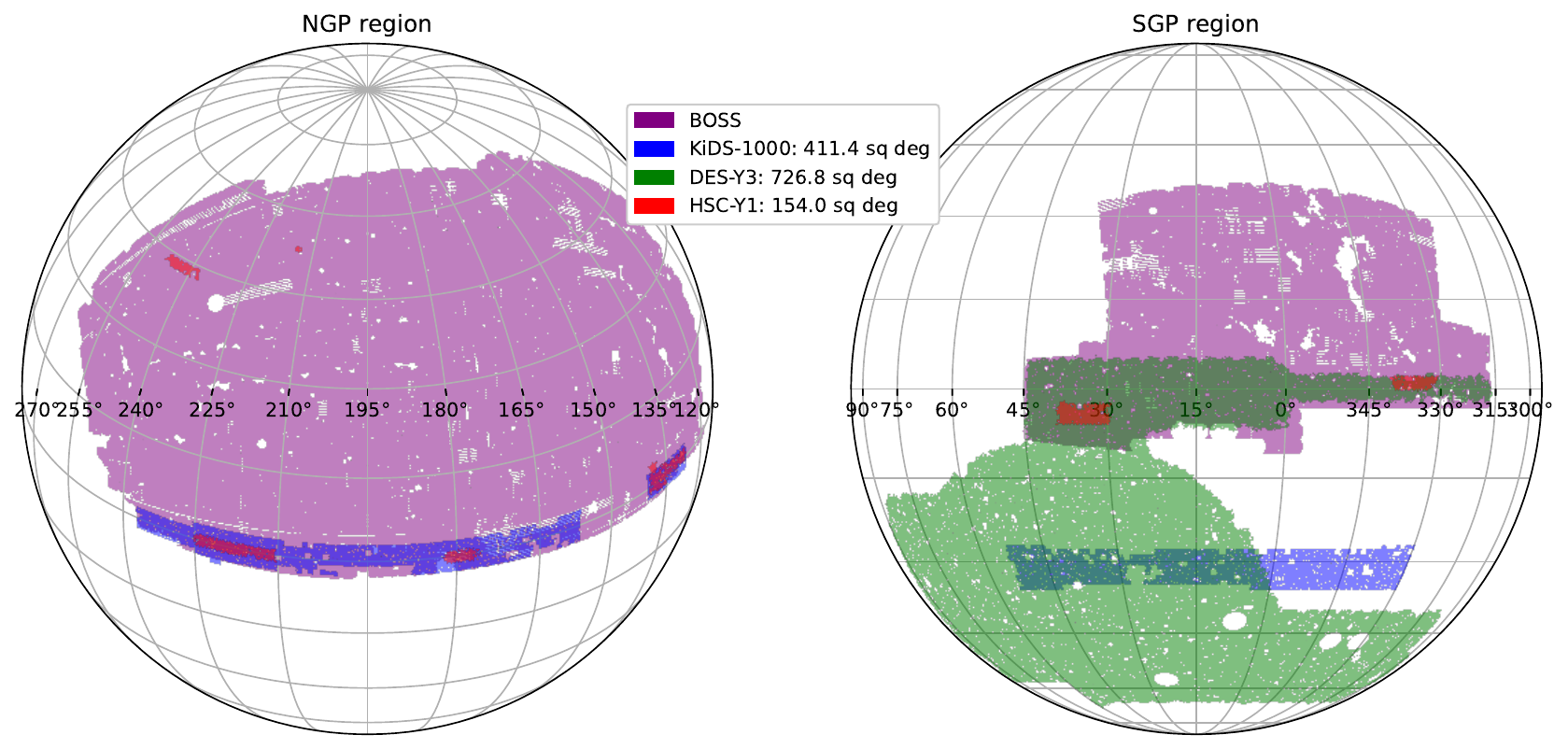}
  \caption{The footprints on the sky of weak lensing surveys: KiDS-1000 (blue), DES-Y3 (green), and HSC-Y1 (red), and their overlap with the BOSS dataset (purple), which is estimated for each survey as 411.4, 726.8, and 154.0 deg$^2$, respectively. The two hemispheres are centred on the North Galactic Pole (NGP) and South Galactic Pole (SGP) regions.}
  \label{fig:map}
\end{figure*}

The Baryon Oscillation Spectroscopic Survey \citep[BOSS,][]{BOSS_1} is a galaxy redshift survey that was performed using the 2.5-m aperture Sloan Telescope at the Apache Point Observatory between 2009 and 2014, as part of the SDSS-III program.  BOSS mapped the distribution of 1.5 million Luminous Red Galaxies (LRG) and quasars across an approximate area of 10{,}000 deg$^2$.  The BOSS collaboration has presented cosmological analysis of this dataset, including the measurement of baryon acoustic oscillation and redshift-space distortions in the galaxy clustering pattern \citep{BOSS_2}.  For our analysis we used the final (Data Release 12) large-scale structure catalogues \citep{BOSS_3}, which combined LOWZ and CMASS LRG samples. In this project, we define 5 lens bins with bin limits $z = [0.2, 0.3, 0.4, 0.5, 0.6, 0.7]$, following \cite{Blake_C}. The redshift distributions of each lens sample can be seen in Figure \ref{fig:nz_lens}. We also utilised unclustered random catalogues provided by the BOSS collaboration, which match the survey selection function.  We combined the overlapping data from BOSS and weak lensing surveys described below to calculate galaxy-galaxy lensing measurements. The sky distributions of BOSS and the three weak lensing surveys, illustrating these overlaps, can be seen in Figure \ref{fig:map}.

\begin{figure}
  \includegraphics[width=\linewidth]{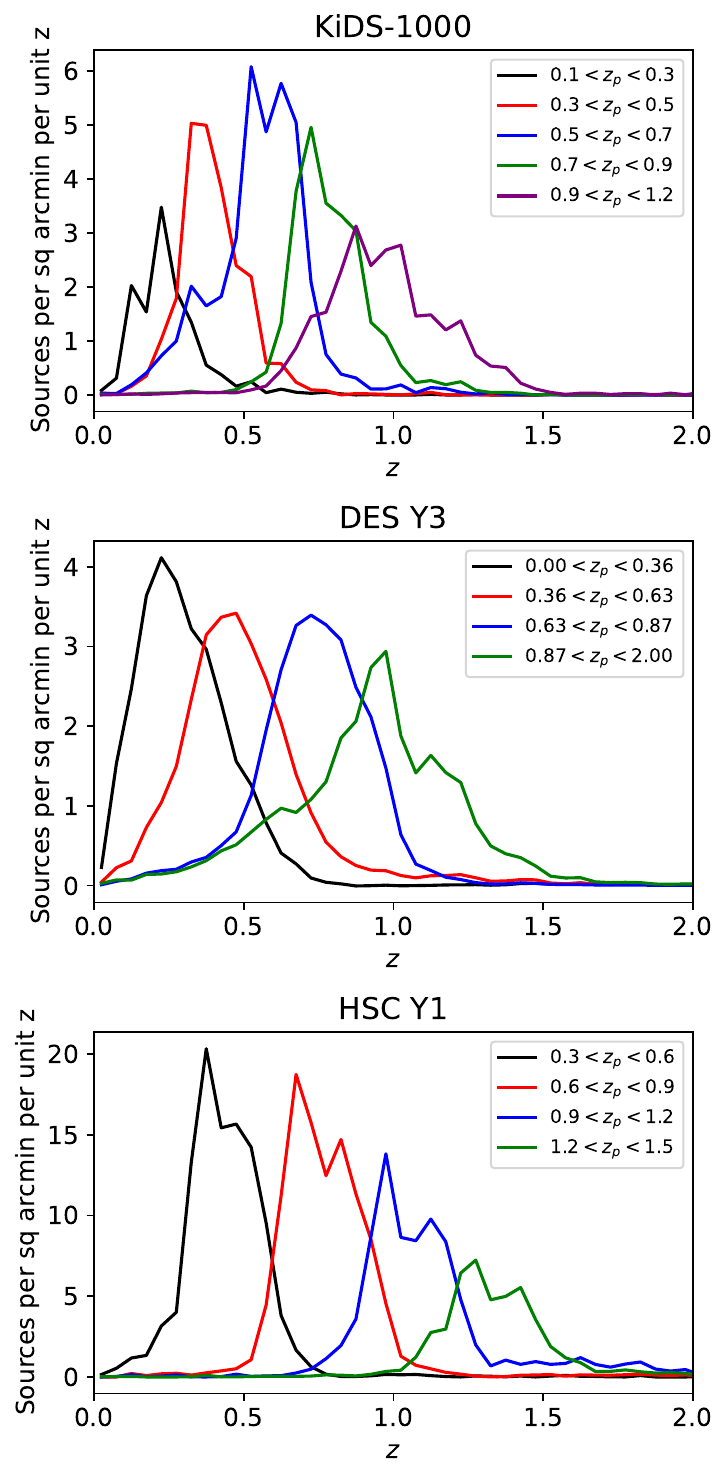}
  \caption{The source redshift distribution of the different tomographic samples of weak lensing surveys, displayed in units of sources per square arcminute per unit redshift.  KiDS-1000 is divided into five tomographic source bins selected in photo-$z$ range $0.1 < z_p < 1.2$, and both DES-Y3 and HSC-Y1 use four tomographic source bins.  The photometric redshift range of DES-Y3 and HSC are $0.0 < z_p < 2.0$ and $0.3 < z_p < 1.5$, respectively.}
  \label{fig:nz_s}
\end{figure}

\subsection{KiDS-1000}
The Kilo-Degree Survey project is a large optical wide-field imaging survey which is optimised for weak gravitational lensing analyses. KiDS was performed with the OmegaCAM instrument on the VLT (Very Large Telescope) Survey Telescope at the European Southern Observatory's Paranal Observatory. The survey uses $u,g,r,i$ optical filters and covers two principal regions of sky, KiDS-North and KiDS South, with a planned final sky coverage of $1{,}350$ deg$^2$. The VISTA-VIKING surveys provide complementary imaging in $Z,Y,J,H,K_s$ near-infrared bands, resulting in deep, wide, nine-band imaging data sets with improved photometric redshifts \citep{2021A&A...647A.124H}.  In this work we used the KiDS-1000 weak lensing catalogue \citep{2021A&A...645A.105G} from the fourth public data release of the project \citep{2019A&A...625A...2K}, where the source shapes were obtained using the {\it lens}fit model-fitting method, and which represents 1006 deg$^2$ of lensing data.  The effective number density of KiDS-1000 is 6.2 arcmin$^{-2}$ and 411.4 deg$^2$ overlaps the BOSS footprint.  We used the 5 source tomographic bins defined by KiDS in our analysis, with photo-$z$ bin limits $z_B = [0.1, 0.3, 0.5, 0.7, 0.9, 1.2]$. The redshift distribution of KiDS-1000 sources can be seen in Figure \ref{fig:nz_s} \citep{2021A&A...647A.124H}. The KiDS collaboration has performed cosmological model fits to the shear correlation functions measured in these tomographic bins \citep{kids1000_cut2}. 

\subsection{DES-Y3}
The largest current weak lensing survey has been performed by the Dark Energy Survey (DES). The DES instrument is installed at the 4-metre Blanco Telescope and uses the Dark Energy Camera \citep{DES_2} and five broad-band filters ($g,r,i,z,Y$).  For this analysis we used DES data taken during the first three years of operations (Y3) between 2013 and 2016, which covers $4{,}143$ deg$^2$ of sky after masking \citep*{2021MNRAS.504.4312G}.  The shape catalogue was created using the self-calibrating shear measurement pipeline {\small METACALIBRATION} \citep{Metacalibration1,Metacalibration2} and contains over $10^8$ objects.  DES-Y3 has an effective source density of 5.6 arcmin$^{-2}$ and overlaps 726.8 deg$^2$ of the BOSS area.  We used four tomographic bins of sources from DES-Y3 with photo-$z$ bin limits $[0.00, 0.36, 0.63, 0.89, 2.00]$, as described in \citet*{DES_nz}. The redshift distribution of DES-Y3 sources is displayed in Figure \ref{fig:nz_s}. The DES collaboration has presented cosmological fits and calibration tests of the shape catalogue \citep*{2022PhRvD.105b3520A, 2022PhRvD.105b3514A, DES_cut}.

\subsection{HSC-Y1}
The Subaru Hyper Suprime-Cam (HSC) survey is currently the deepest large-scale weak lensing survey. HSC operates at the Subaru 8.2m telescope and the HSC Strategic Program aims to cover 1400 deg$^2$ of the sky using $g, r, i, z, Y$ filters.  In our analysis we use the first year data release of the HSC weak lensing catalogue \citep{2018PASJ...70S..25M} that has an effective source number density of 17 arcmin$^{-2}$, with shapes measured using a re-Gaussianization PSF correction method, and covers an overlap area of 154.0 deg$^2$ with BOSS. For this work we use four tomographic bins of sources with photo-$z$ bin limits $[0.3, 0.6, 0.9, 1.2, 1.5]$. The redshift distribution of HSC-Y1 sources in each tomographic bin can be seen in Figure \ref{fig:nz_s}. Cosmological analyses of this shape catalogue have been presented by \cite{2019PASJ...71...43H} and \cite{hsc_cut}.

\section{Shear Ratio Measurement}
\label{sec:measurements}

\subsection{Average Tangential Shear}
\label{sec:gammat}

Galaxy-galaxy lensing describes the distortion of background galaxy images in the presence of foreground dark matter halos occupied by the lens galaxies. This distortion stretches the background sources perpendicular to the line joining the source and lens galaxies on the sky. The magnitude of this tangential shear, $\gamma_\mathrm{t}(\theta)$, and the related excess surface density of matter around the lens at projected scale $R$, $\Delta\Sigma(R)$, provide a means of learning about the local dark matter profile and galaxy environment.

In this work we use $\gamma_\mathrm{t}(\theta)$ for performing a shear ratio measurement.  We used the estimator of the average tangential shear,
\begin{equation}\label{gammat}
    \hat{\gamma}_\mathrm{t}(\theta) = \frac{\sum\limits_{\mathrm{ls}} w_\mathrm{l} w_\mathrm{s} \, e_\mathrm{{t,ls}} - \sum\limits_\mathrm{{rs}} w_\mathrm{r} w_\mathrm{s} \, e_\mathrm{{t,rs}} }{\sum\limits_\mathrm{{rs}} w_\mathrm{r} w_\mathrm{s}}
\end{equation}
\cite[see for example,][]{Blake_C}.  This estimator combines catalogues of sources ($\mathrm{s}$), lenses ($\mathrm{l}$) and an unclustered lens catalogue ($\mathrm{r}$) with the same selection function as the lenses, where $w$ is the weight of each object and $e_\mathrm{t}$ is the tangential ellipticity of the source, projected on an axis normal to the line joining the source and lens or random lens. Equation~\ref{gammat} assumes that the weights are normalised such that $\sum_\mathrm{l} w_\mathrm{l} = \sum_\mathrm{r} w_\mathrm{r}$.  We measured the average tangential shear in 10 logarithmically-spaced separation bins in the range $0.03 < \theta < 3$ degrees.

We computed the terms in Equation \ref{gammat} from the source and lens catalogues using the {\small TREECORR} correlation function code\footnote{\url{https://github.com/rmjarvis/TreeCorr}} \citep{2004MNRAS.352..338J}. The first term in the estimator implicitly includes the ``boost factor'',
\begin{equation}\label{boost}
B(\theta) = \frac{\sum\limits_{\mathrm{ls}} w_\mathrm{l} w_\mathrm{s}}{\sum\limits_{\mathrm{rs}} w_\mathrm{r} w_\mathrm{s}}, 
\end{equation}
which quantifies physical source-lens associations not caused by gravitational lensing, which can occur in cases where the source and lens redshift distributions partially overlap.

We determined a Gaussian analytical covariance of the $\gamma_\mathrm{t}$ measurements across lens bins, source bins and scales, following the methods described by \cite{Blake_C}.  Our analytical covariance includes the sample variance, noise and mixed contributions.  We assume a fiducial non-linear matter power spectrum, the lens and source distributions displayed in Figures \ref{fig:nz_lens} and \ref{fig:nz_s}, the effective source number densities and shape noise measured by the weak lensing collaborations, and assumed linear galaxy bias factor $b=2$ for the lens galaxies.  The covariance is noise-dominated on small scales, so the assumption of linear bias does not have a significant impact on our results.  We evaluated the covariances for the overlap areas indicated in Figure \ref{fig:map}.

\subsection{Shear Ratio}
\label{sec:shearratio}

We can use galaxy-galaxy lensing measurements to perform an SR test. We start with a simple illustration of why the SR is a useful probe of cosmology.  In general, $\gamma_\mathrm{t}(\theta)$ for a single lens-source pair can be expressed as,
\begin{equation}\label{eq_7}
\gamma_\mathrm{t}(\theta) = \frac{\Delta\Sigma(R = \theta D_\mathrm{l})}{\Sigma_\mathrm{crit}(z_\mathrm{l},z_\mathrm{s})}
\end{equation}
where $\Sigma_\mathrm{crit}$ is a geometrical factor depending on the distances of the lens galaxy, $D_\mathrm{l}$, and source galaxy, $D_\mathrm{s}$, and the source-lens distance $D_\mathrm{ls}$, according to,
\begin{equation}\label{eq_8}
\Sigma_\mathrm{crit}^{-1}(z_\mathrm{l},z_\mathrm{s}) = \frac{4\pi G}{c^2}\frac{D_\mathrm{ls} D_\mathrm{l}}{D_\mathrm{s}}
\end{equation}
Equation \ref{eq_7} can be used to form a ratio of two tangential shear measurements sharing the same lens with two different sources $\mathrm{s_i}$ and $\mathrm{s_j}$, where $\Delta\Sigma$ cancels out since it is a property of the lens only. The shear ratio in this case is given by,
\begin{equation}\label{eq_9}
r^{(\mathrm{l,s_i,s_j})}(\theta) = \frac{\gamma^{\mathrm{l,s_i}}_\mathrm{t}(\theta)}{\gamma^{\mathrm{l,s_j}}_t(\theta)} = \frac{\Sigma_\mathrm{crit}^{-1}(z_\mathrm{l},z_\mathrm{si})}{\Sigma_\mathrm{crit}^{-1}(z_\mathrm{l},z_\mathrm{sj})}
\end{equation}
Hence, the ratio of $\gamma_\mathrm{t}$ just depends on the distances of the lens and source galaxies, not on the angular separation $\theta$ or nature of the lens galaxy, and is therefore a measure of the cosmological distance-redshift relation independently of any astrophysical properties of the lens galaxy \citep{Jain&Taylor}.

We now introduce some additional complexities to the model.  We do not analyse a single lens and source, but rather a distribution over redshift.  We write the lens redshift distribution as $n_\mathrm{l}(z)$, and a sample of source galaxies as $n_\mathrm{s}(z)$. In this case, the $\Sigma_\mathrm{crit}$ factor in Equation \ref{eq_9} can be expressed as,
\begin{equation}\label{eq_10}
\Sigma^{-1 \; \mathrm{i,j}}_\mathrm{crit,eff} = \int^{z^{\max}_\mathrm{l}}_{0} dz_\mathrm{l} \int^{z^{\max}_\mathrm{s}}_{0} dz_\mathrm{s} \, n^\mathrm{i}_\mathrm{l}(z_\mathrm{l}) \, n^\mathrm{j}_\mathrm{s}(z_\mathrm{s}) \, \Sigma^{-1}_\mathrm{crit}(z_\mathrm{l},z_\mathrm{s})
\end{equation}
\citep*[e.g.,][]{2022PhRvD.105h3529S}, where $\mathrm{i}$ and $\mathrm{j}$ are the indices of the lens and source bins, respectively. Hence the ratio can be expressed by the revised equation,
\begin{equation}\label{eq_11}
r^{(\mathrm{l,s_i,s_j})}(\theta) = \frac{\Sigma_\mathrm{crit,eff}^{-1 \; \mathrm{l,s_i}}}{\Sigma_\mathrm{crit,eff}^{-1 \; \mathrm{l,s_j}}}
\end{equation}
Equation \ref{eq_11} is valid for the case of a lens redshift distribution that is narrow enough, neglecting other effects on the tangential shear.  In Section \ref{sec:shearmodel} we describe the full model we adopt for the average tangential shear, including other astrophysical effects such as intrinsic alignments and magnification, which allows us to use the shear ratio to probe a wider range of astrophysical parameters.

Since Equation \ref{eq_11} is independent of source-lens angular separation $\theta$ (in a first-order approximation), we can increase the accuracy of determining the shear ratio by averaging the measurement over separations,
\begin{equation}
r^{(\mathrm{l,s_i,s_j})} \equiv \left \langle \frac{\gamma^{\mathrm{l,s_i}}_\mathrm{t}(\theta)}{\gamma^{\mathrm{l,s_j}}_\mathrm{t}(\theta)} \right \rangle_\theta = \left \langle r^{(\mathrm{l,s_i,s_j})}(\theta) \right \rangle_\theta
\label{eq:averagesr}
\end{equation}
We will express our final shear ratio measurements in this angle-averaged form.  When averaging over scales, we follow \citet*{2022PhRvD.105h3529S} and combine a range $2-6 \, h^{-1}$ Mpc (as converted into angular separations for each lens redshift bin).  As discussed by \citet*{2022PhRvD.105h3529S}, this cut is motivated by (on large scales) the consideration of adding $\gamma_\mathrm{t}(\theta)$ to the data vector for scales $> 6 \, h^{-1}$ Mpc without double-counting information, and (on small scales) being cautious about extending the fit into the ``1-halo'' regime, as tested using simulations. When converted to angular separations using $\theta = R/D_\mathrm{l}$, our fitting range corresponds to: $[9.72-29.15']$, $[7.12-21.35']$, $[5.68-17.03']$, $[4.77-14.3']$ and $[4.14-12.42']$ in the five lens bins.  When constructing our set of shear ratios across the source tomographic samples, we always use the highest-redshift source bin as the denominator of Equation~\ref{eq:averagesr}, since this measurement contains the highest signal, and ratio the measurements for the other source tomographic bins to this reference.  The ratios amongst the other tomographic bins contain no additional information.

\subsection{Bayesian method for SR measurement}

Evaluating the SR from measurements of the average tangential shear involves forming the ratio of two noisy quantities.  Even if these two quantities are unbiased estimates with Gaussian error distributions, their ratio will be a biased measurement ($\langle x/y \rangle \ne \langle x \rangle/\langle y \rangle$) with a non-Gaussian error distribution.  In this section we describe our Bayesian method for determining the best-fitting ratio and its error, inspired by \cite{bayes_fit}. 

Suppose the two sets of $\gamma_\mathrm{t}$ measurements as a function of separation are data vectors $\mathbf{x}_1$ and $\mathbf{x}_2$, and we want to determine the posterior probability distribution of the ratio $r = \langle \mathbf{x}_1/\mathbf{x}_2 \rangle$.  The data vectors $\mathbf{x}_1$ and $\mathbf{x}_2$ are correlated, and we describe this in terms of the concatenated data vector $\mathbf{x} = (\mathbf{x}_1,\mathbf{x}_2)$ which has covariance $\mathbf{C}$.  Hence, we seek to determine $P(r|\mathbf{x})$, given this covariance.

We use the $\gamma_\mathrm{t}$ model predictions $\mathbf{m}_1$ and $\mathbf{m}_2$ to create the data model,
\begin{equation}
\begin{split}
\mathbf{x}_1 &= a \, \mathbf{m}_1 + \mathbf{n}_1 \\
\mathbf{x}_2 &= a \, \mathbf{m}_2 + \mathbf{n}_2
\end{split}
\label{bayesian_2}
\end{equation}
where $a$ is a constant amplitude we marginalise over, and $\mathbf{n}_1$ and $\mathbf{n}_2$ are noise vectors.  We assume $\mathbf{m}_1 = r \, \mathbf{m}_2$, where $r$ is the shear ratio we are fitting.  The probability $P(r|\mathbf{x})$ can then be evaluated using Bayes' Theorem as,
\begin{equation}
\begin{split}
\mathrm{P}(r|\mathbf{x}) &\propto \int da \, \mathrm{P}(\mathbf{x}|r,a) \\
&= \int da \, \exp \bigg[-\frac{1}{2} (\mathbf{x}^T -a \mathbf{m}^T) \mathbf{C}^{-1} (\mathbf{x} - a \mathbf{m})\bigg] \\
&= \int da \, \exp \bigg[-\frac{1}{2} ( \mathbf{x}^T \mathbf{C}^{-1} \mathbf{x} -a \mathbf{x}^T \mathbf{C}^{-1} \mathbf{m} \\
& \hspace{1cm} - a \mathbf{m}^T \mathbf{C}^{-1} \mathbf{x} + a^2 \mathbf{m}^T \mathbf{C}^{-1} \mathbf{m}) \bigg]
\end{split}
\label{bayesian_5}
\end{equation}
where $\mathbf{m} = (\mathbf{m}_1,\mathbf{m}_2)$ is the concatenated model vector. Since $\mathbf{x}^T \mathbf{C}^{-1} \mathbf{x}$ is independent of $a$ and $r$, it can be absorbed into an overall normalisation constant. Using the fact that $\mathbf{x}^T \mathbf{C}^{-1} \mathbf{m} = \mathbf{m}^T \mathbf{C}^{-1} \mathbf{x}$, Equation \ref{bayesian_5} becomes,
\begin{equation}
\begin{split}
\mathrm{P}(r|\mathbf{x}) &\propto \int da \, \exp \bigg[-\frac{1}{2} a^2 \mathbf{m}^T \mathbf{C}^{-1} \mathbf{m} +a \mathbf{x}^T \mathbf{C}^{-1} \mathbf{m}  \bigg] \\
&= \int da \, \exp (-\lambda^2 a^2 + \mu a)
\end{split}
\label{bayesian_7}
\end{equation}
where $\lambda^2 = \frac{1}{2} \mathbf{m}^T \mathbf{C}^{-1} \mathbf{m}$ and $\mu = \mathbf{x}^T \mathbf{C}^{-1} \mathbf{m}$. This relation can be re-written as,
\begin{equation}\label{bayesian_8}
\mathrm{P}(r|\mathbf{x}) \propto \int da \exp \bigg[ -\bigg( \lambda a - \frac{\mu}{2\lambda} \bigg)^2 + \frac{\mu^2}{4 \lambda^2}\bigg]
\end{equation}
which can readily be integrated to give,
\begin{equation}\label{bayesian_9}
\mathrm{P}(r|\mathbf{x}) \propto \frac{1}{\lambda} \exp \bigg(\frac{\mu^2}{4 \lambda^2} \bigg).
\end{equation}
Our method for finding $P(r|\mathbf{x})$ is to vary $r$ over a discrete grid  and, for each value of $r$, evaluate $\lambda^2 = \frac{1}{2} \mathbf{m}^T \mathbf{C}^{-1} \mathbf{m}$ and $\mu = \mathbf{x}^T \mathbf{C}^{-1} \mathbf{m}$, after setting $\mathbf{m}_1 = r \mathbf{m}_2$.  We then evaluate the shear ratio and its error as the mean and standard deviation of this probability distribution.  We further generalise this method to determine the covariance between two different shear ratio measurements.  We note that this methodology can be applied in a single angular separation bin to measure $r^{(\mathrm{l,s_i,s_j})}(\theta)$, or across a set of angular bins, to measure $r^{(\mathrm{l,s_i,s_j})} = \left \langle r^{(\mathrm{l,s_i,s_j})}(\theta) \right \rangle_\theta$.

\begin{figure*}
  \includegraphics[width=\linewidth]{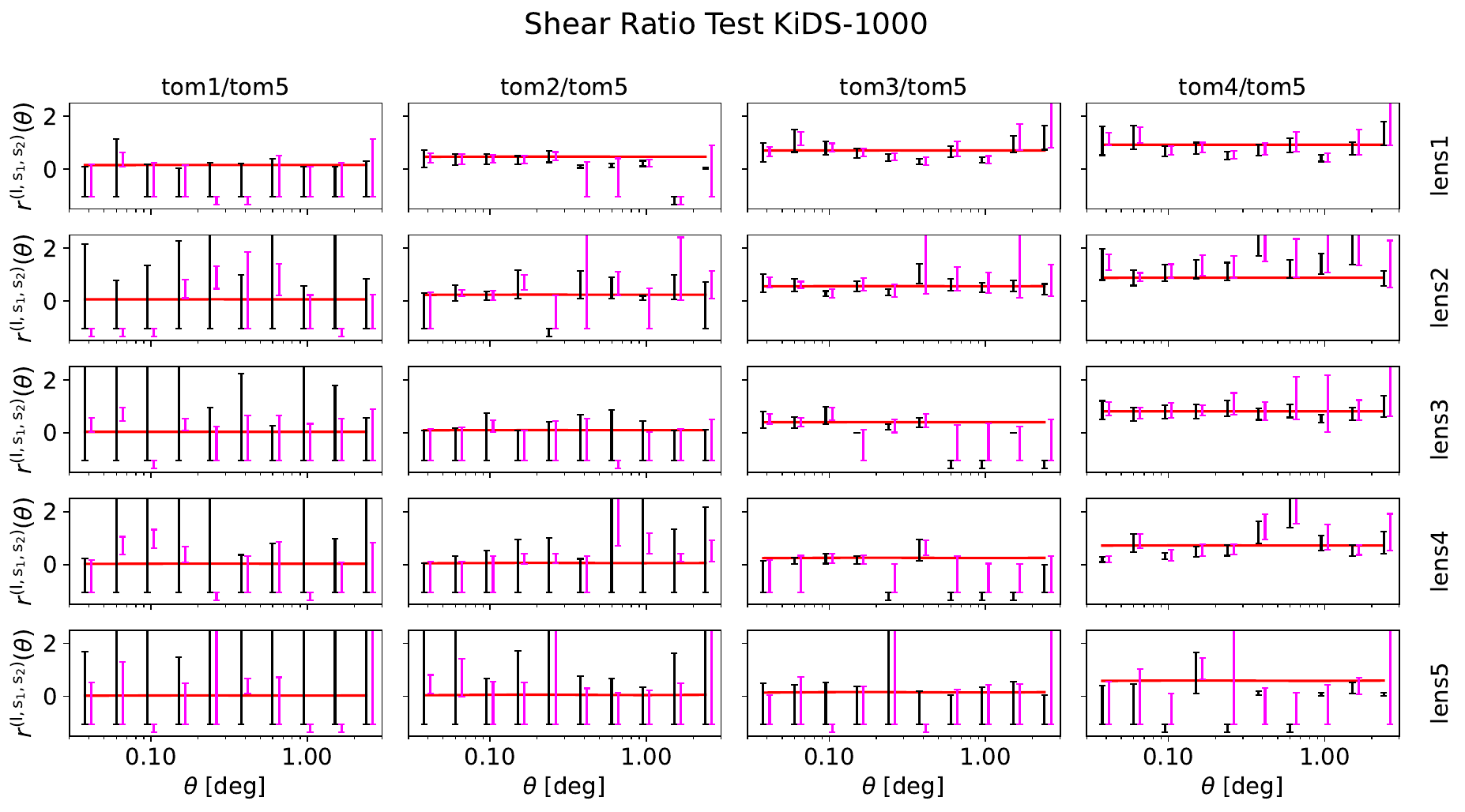}
  \caption{The shear ratio as a function of angular separation for KiDS-1000 sources relative to BOSS lenses.  The columns display shear ratios for different source tomographic bins with reference to the highest-redshift source bin, and the rows correspond to the five different BOSS lens samples.  The black measurements show the result of the direct evaluation of the shear ratio, and the pink measurements adopt the Bayesian fitting method. The solid red line indicates our fiducial model for the SR, which is computed from the $\gamma_t(\theta)$ models using the fiducial parameter settings}.
  \label{fig:SR_kids}
\vspace{0.5cm}
\end{figure*}

\begin{figure}
  \includegraphics[width=\linewidth]{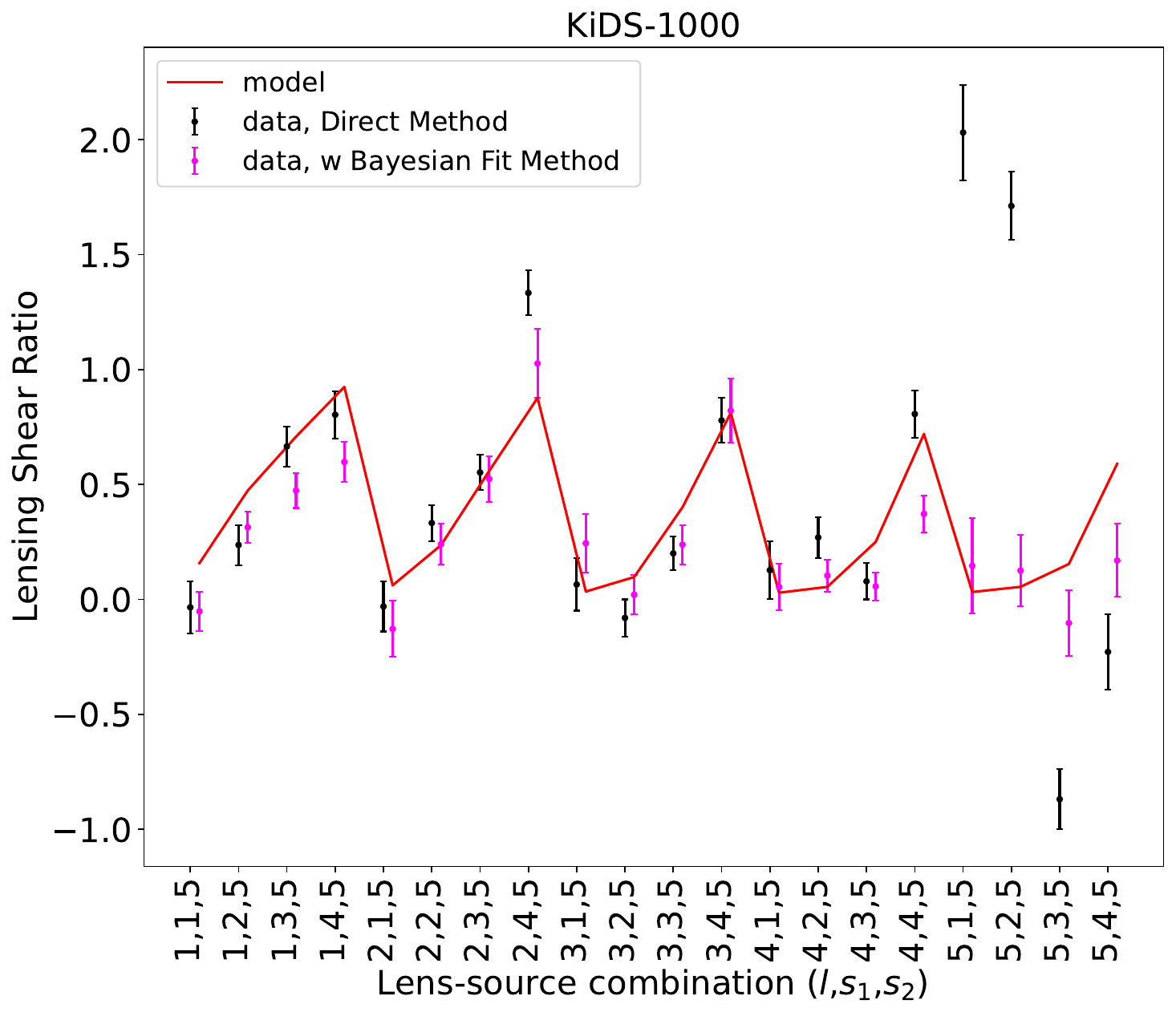}
  \caption{Comparison between the angle-averaged shear ratio measured using the direct method (black data points) and the Bayesian fitting method (pink data points), for the KiDS-1000 sources relative to BOSS lenses.  We do not apply any scale cuts. The red solid line is the theoretical model calculated using the fiducial parameters.  The series of measurements represent different combinations of lens and source bins, as indicated by the $x$-axis label.  When we directly divide two $\gamma_\mathrm{t}$ measurements, if the denominator is close to zero the result of the ratio will be biased and possess non-Gaussian errors.  The Bayesian fitting method addresses this problem by constructing the full posterior probability distribution for the ratio, given the data and covariance.}
  \label{fig:SR_KiDS_2}
\end{figure}

To illustrate the effects of this Bayesian methodology, we compute the SR as a function of angular separation for each weak lensing survey using a direct method (taking a simple ratio of two $\gamma_\mathrm{t}$ measurements) and the Bayesian fitting method described above. The comparison of these two measurements can be seen in Figure \ref{fig:SR_kids} (using KiDS-1000 as an example, where the results for the other weak lensing surveys have a consistent appearance).  In Figure \ref{fig:SR_KiDS_2} we compare these measurement techniques as applied to the angle-averaged shear ratio (we describe the models shown in the figures in Section \ref{sec:shearmodel} below).  In both situations, the direct ratio method appears to contain significant biases in cases where the denominator of the shear ratio is close to zero (which can arise, for example, when a significant portion of the source distribution is located in front of the lenses, which can happen for high-redshift lenses).

\begin{figure}
  \includegraphics[width=\linewidth]{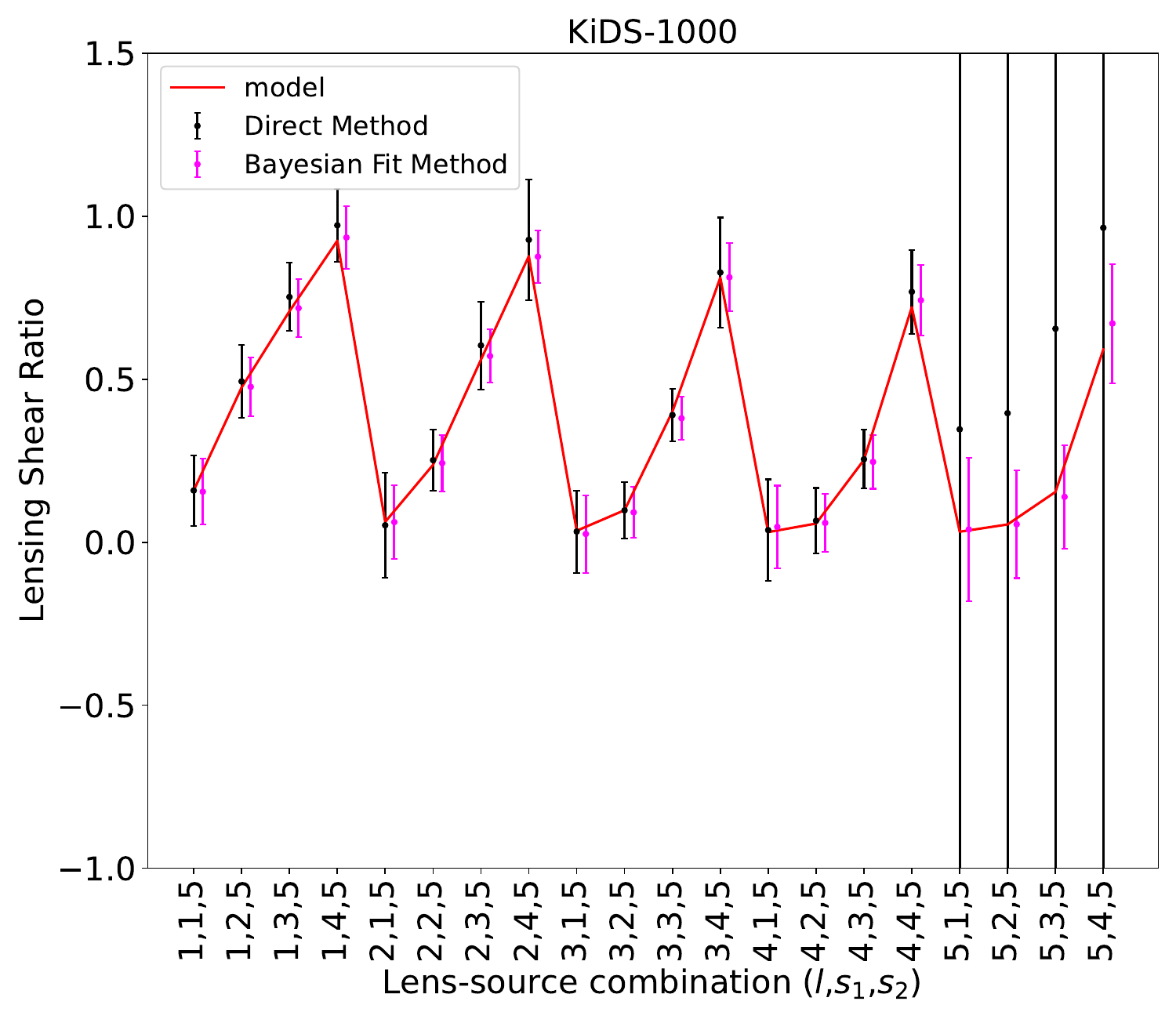}
  \caption{The mean and standard deviation of Monte Carlo realisations of measurements of the angle-averaged shear ratio for the KiDS-1000 sources relative to BOSS lenses.  The realisations are generated by applying correlated errors to the fiducial $\gamma_\mathrm{t}$ models, derived from the analytical covariance.  We find that the variance of the direct method is significantly larger than the Bayesian method in the final lens redshift bin, consistent with the excess scatter visible in Figure \ref{fig:SR_KiDS_2}.}
  \label{fig:monte_kids}
\vspace{0.5cm}
\end{figure}

We created a set of Monte Carlo simulations to validate the performance of the Bayesian SR measurement method compared to the direct method.  Using KiDS-1000 as an example, we sampled 100 realizations of noisy $\gamma_\mathrm{t}$ data sets using the fiducial model and a Cholesky decomposition of the analytical covariance.  We measured the angle-averaged SR for each realization, and hence determined the mean and standard deviation of the measurements across the realisations, using the two methods (direct and Bayesian). The result of this experiment can be see in Figure \ref{fig:monte_kids}.  Whilst the two sets of measurements agree well for the first four lens redshift bins, the direct estimator applied to the final lens redshift bin suffers a large variance which can be attributed to the non-Gaussian error of a ratio, as discussed by \cite{bayes_fit}, whereas the variance of the Bayesian estimate is smaller in this case.  This behaviour is consistent with the significant scatter in the direct measurements for this lens bin for the data, visible in Figure \ref{fig:SR_KiDS_2}.

\begin{figure}
  \includegraphics[width=\linewidth]{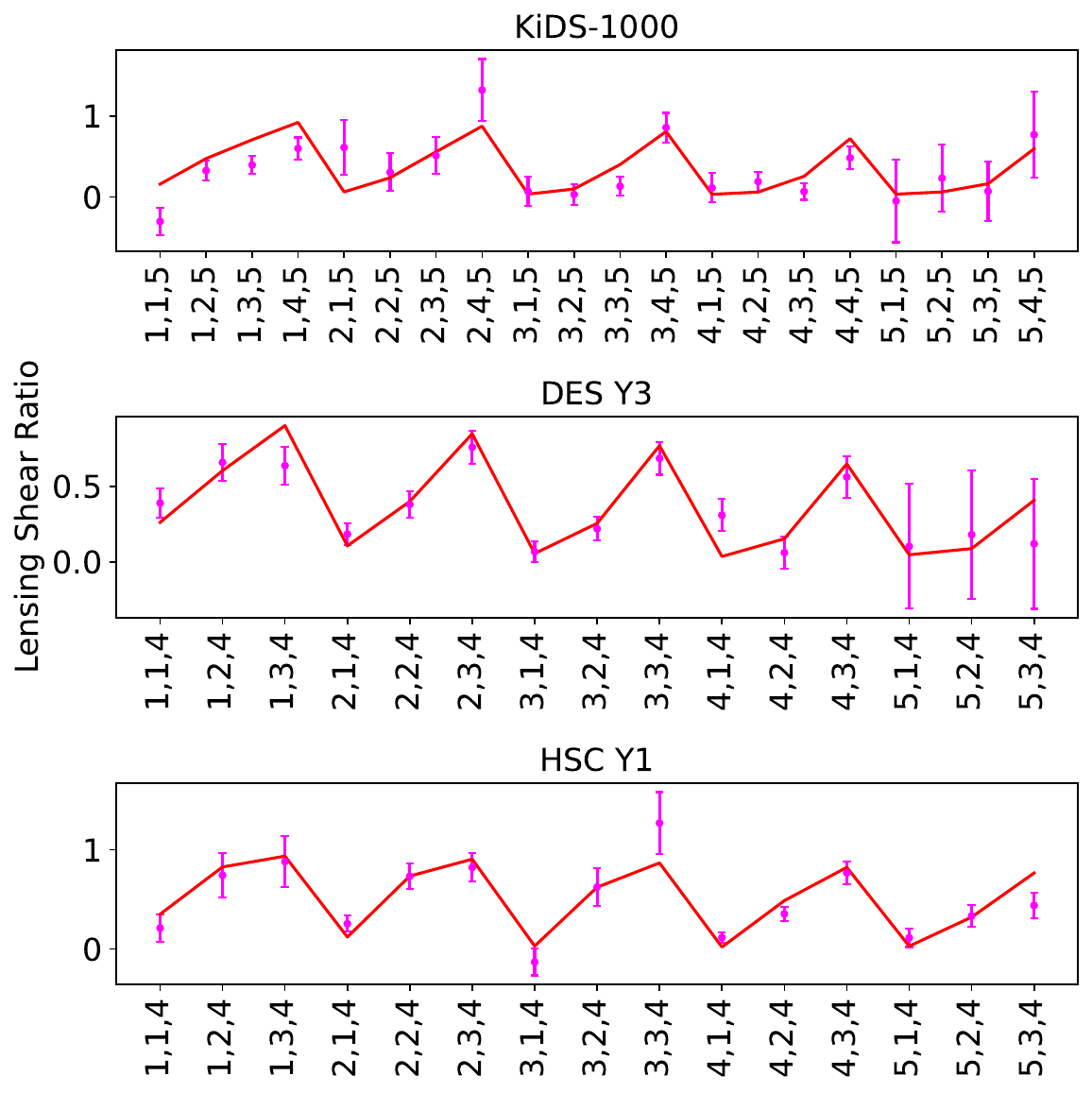}
  \caption{The angle-averaged shear ratio measurements for KiDS-1000, DES-Y3, and HSC-Y1 sources relative to BOSS lenses, determined using the Bayesian fitting method, and compared to the fiducial models in each case.  We averaged the shear ratio measurements over scales $2-6 \, h^{-1}$ Mpc, following \citet*{2022PhRvD.105h3529S}.  The series of measurements in each panel represent different combinations of lens bins $(l)$ and source bins $(s_1, s_2)$, as indicated by the $x$-axis label in the combination $(l, s_1, s_2)$.  These measurements are used as inputs to our cosmological and astrophysical parameter fits.}
  \label{fig:SR_all}
\end{figure}

Our results for the angle-averaged shear ratio for the three weak lensing surveys are displayed in Figure \ref{fig:SR_all}, using the Bayesian fitting method. We will use these results as inputs to the cosmological and astrophysical parameter fits, which are performed using the models described in the next section.

\section{Model}
\label{sec:shearmodel}

\subsection{Average Tangential Shear Model}

We adopt a full model for the average tangential shear as used in the DES-Y3 $3\times2$-pt and shear ratio analyses \citep*{2022PhRvD.105b3520A,2022PhRvD.105h3529S}, which includes explicit modeling of effects such as uncertainty of source redshift, lens magnification, intrinsic alignments (IA) and multiplicative shear biases.  We refer to \citet*{2022PhRvD.105h3529S} for a full description of the model, and in this section we briefly summarise its main terms.

The average tangential shear for a lens sample $\mathrm{i}$ and source sample $\mathrm{j}$ is evaluated as,
\begin{equation}\label{eq_13}
\gamma^{\mathrm{i,j}}_\mathrm{t}(\theta) = (1+m^\mathrm{j}) \sum \limits_{\ell} \frac{2\ell+1}{4\pi \ell(\ell+1)} \overline{P^2_{\ell}}(\theta_{\min},\theta_{\max}) C^{\mathrm{ij}}_{\mathrm{gm},\mathrm{tot}}(\ell)
\end{equation}
where $\ell$ is the 2D multipole moment, $\overline{P^2_{\ell}}(\theta_{\min},\theta_{\max})$ is the bin-averaged associated Legendre  polynomial within an angular bin with limits $(\theta_{\min},\theta_{\max})$, $m^\mathrm{j}$ is the multiplicative uncertainty on the shape measurements, and $C^{\mathrm{ij}}_{\mathrm{gm},\mathrm{tot}}(\ell)$ is the total angular galaxy-matter cross-spectrum that includes IA, lens magnification and cross-terms between two effects:
\begin{equation}\label{eq_14}
C^{\mathrm{ij}}_{\mathrm{gm},\mathrm{tot}}=C^{\mathrm{ij}}_{\mathrm{gm}}+C^{\mathrm{ij}}_{\mathrm{gm},\mathrm{IA}}+C^{\mathrm{ij}}_{\mathrm{gm},\mathrm{mag}}+C^{\mathrm{ij}}_{\mathrm{gm},\mathrm{IA} \times \mathrm{mag}}
\end{equation}
The key model ingredients we are testing are the source redshift distributions, intrinsic alignments, and cosmological parameters, especially $S_8$.

Having calculated the tangential shear model for a given parameter set, we can determine the prediction for the SR for the lens redshift bin $\mathrm{i}$ between source redshift bins $\mathrm{j}$ and $\mathrm{k}$, averaging over angular separations using Equation~\ref{eq:averagesr}, where the tangential shear correlation function for each angular bin can be expressed as a transformation of the galaxy-matter angular cross-power spectrum $C_{\mathrm{gm},\mathrm{tot}}(\ell)$ using Equation \ref{eq_13}.

We used the {\small COSMOSIS} software \citep{2015A&C....12...45Z} to evaluate these model predictions.  The primary angular cross-power spectrum (the first term in Equation \ref{eq_14}) can be written as the projection of the 3D galaxy-matter cross-power spectrum $P_{\mathrm{gm}}(k,z)$ using Limber’s approximation \citep{Limber, Limber2} and assuming a flat Universe cosmology:
\begin{equation}\label{eq_15}
C^{\mathrm{ij}}_{\mathrm{gm}}(\ell) = \frac{3H_0^2\Omega_\mathrm{m}}{2c^2}\int d\chi N^\mathrm{i}_\mathrm{l}(\chi) \frac{g^\mathrm{j}(\chi)}{a(\chi)\chi} P_{\mathrm{gm}} \left ( \frac{\ell+1/2}{\chi},z(\chi) \right) 
\end{equation}
where $H_0$ is the Hubble constant, $c$ is the speed of light, and the distance distribution of the lenses is given by,
\begin{equation}\label{eq_16}
N^\mathrm{i}_\mathrm{l}(\chi) = \frac{n^\mathrm{i}_\mathrm{l} (z)}{\overline{n}^\mathrm{i}_\mathrm{l}} \frac{dz}{d\chi} 
\end{equation}
In Equation \ref{eq_16}, $\chi$ is the comoving distance to redshift $z$, $a$ is the scale factor, $n^\mathrm{i}_\mathrm{l}$ is the lens redshift distribution as displayed in Fig.\ref{fig:nz_s}, $\overline{n}^\mathrm{i}_\mathrm{l}$ is the mean number density of the lens galaxies, and $g^\mathrm{j}(\chi)$ is the lensing efficiency kernel:
\begin{equation}\label{eq_17}
g(\chi) = \int^{\chi_{\lim}}_{\chi} d\chi ' N^\mathrm{j}_\mathrm{s}(\chi ') \left( \frac{\chi '-\chi}{\chi '} \right)
\end{equation}
where $\chi_{\lim}$ is the limiting comoving distance of the source galaxy sample. The source distribution is given analogously to Equation~\ref{eq_16},
\begin{equation}
N^\mathrm{i}_\mathrm{s}(\chi) = \frac{n^\mathrm{i}_\mathrm{s} (z -\Delta z^\mathrm{i})}{\overline{n}^\mathrm{i}_\mathrm{s}} \frac{dz}{d\chi} 
\end{equation}
where $\Delta z^\mathrm{i}$ is the shift in the mean redshift of the source redshift distributions shown in Figure~\ref{fig:nz_s}, which we have introduced as a free parameter.

The galaxy-matter cross-power spectrum in Equation \ref{eq_15} is given by,
\begin{equation}\label{eq_18}
P^{\mathrm{ij}}_{\mathrm{gm}} = b^\mathrm{i} P^{\mathrm{ij}}_{\mathrm{mm}}
\end{equation}
where $b$ is the galaxy bias of the lens sample.  In the shear ratio, galaxy bias cancels out and has no significant effect.  The non-linear matter power spectrum $\mathrm{P}_{\mathrm{mm}}$ is computed using the \cite{Takahashi_2012} version of {\small HALOFIT}, and the linear power spectrum with {\small CAMB} \citep{2000ApJ...538..473L}.

Moving to the other angular power spectrum components of Equation \ref{eq_14}, the intrinsic alignment contribution is given by,
\begin{equation}\label{eq_20}
C^{\mathrm{ij}}_{\mathrm{gm},\mathrm{IA}}(\ell) = \int d\chi \frac{N^\mathrm{i}_\mathrm{l}(\chi) N^\mathrm{j}_\mathrm{s}(\chi)}{\chi^2} P_{\mathrm{gI}} \left( k=\frac{\ell+1/2}{\chi},z(\chi)\right)
\end{equation}
where $P_{\mathrm{gI}} = b P_{\mathrm{GI}}$. Intrinsic alignment refers to the correlation of the source galaxy orientation and the underlying large-scale structure due to physical association rather than gravitational lensing. This effect is only present in galaxy-galaxy lensing measurements if the lens and source galaxies overlap in redshift. 

We consider two different models for intrinsic galaxy alignments: NLA (Non-Linear Alignment) \citep{2007Bridle}, which we use for the KiDS-1000 analysis, and TATT (Tidal Alignment and Tidal Torquing) \citep{2019Blazek}, which we apply to DES-Y3 and HSC-Y1.  We chose these different models for each survey in order to compare with the results previously published by these collaborations.

In the NLA model, we use:
\begin{equation}\label{eq_nla}
P_{\mathrm{GI}}(k,z) = a_1(z) P_{\mathrm{mm}}(k) 
\end{equation}
with 
\begin{equation}\label{eq_22}
a_1(z) = -A_1 \overline{C}_1 \frac{\rho_\mathrm{crit} \Omega_\mathrm{m}}{D(z)} \left( \frac{1+z}{1+z_0} \right)^{\eta_1}
\end{equation}
where $D(z)$ is the linear growth factor and $\overline{C}_1$ is a normalisation constant with value $\overline{C}_1 = 5 \times 10^{-14} h^{-2} M_{\odot}^{-1} \text{Mpc}^3$ \citep{2004hirata, 2007Bridle}. The NLA model hence has 2 free parameters, $A_1$ and $\eta_1$.

In the TATT model of intrinsic alignments we use,
\begin{equation}\label{eq_21}
P_{\mathrm{GI}}(k,z) = a_1(z) P_{\mathrm{mm}}(k) + a_{1\delta}(z) P_{0|0E}(k) + a_2(z) P_{0|E2}(k)
\end{equation}
where the full expression for the power spectra of the second and third terms can be found in \cite{2019Blazek}.  We calculate these power spectra using the public code {\small FAST-PT} \citep{fastpt_1, fastpt_2}. The coefficient $a_1$ in the Equation \ref{eq_21} is defined by Equation \ref{eq_22} and the other coefficients are given by,
\begin{equation}\label{eq_23}
a_2(z) = 5A_2 \overline{C}_1 \frac{\rho_\mathrm{crit} \Omega_\mathrm{m}}{D^2(z)} \left( \frac{1+z}{1+z_0} \right)^{\eta_2}
\end{equation}
and
\begin{equation}\label{eq_24}
a_{1\delta}(z) = b_\mathrm{TA} \, a_1(z)
\end{equation}
The TATT model hence has five parameters: $A_1$, $A_2$, $\eta_1$, $\eta_2$ and $b_\mathrm{TA}$ (which accounts for the fact that the shape is preferentially sampled in overdense regions). The $A_1$ and $\eta_1$ parameters match those used in the NLA model.

Another effect of large-scale structure that we need to consider is lens magnification. This is an effect of gravitational lensing that causes a lens to become larger in size whilst having the same surface brightness. The lens magnification angular cross-power spectrum is denoted as $C^{\mathrm{ij}}_{\mathrm{gm, mag}}$ in Equation \ref{eq_14} and can be written as,
\begin{equation}\label{mag}
C^{\mathrm{ij}}_{\mathrm{gm, mag}}(\ell) = 2 \, \left( \alpha^i - 1 \right) \, C^{\mathrm{ij}}_{\mathrm{mm}}(\ell)
\end{equation}
where $\alpha^i$ is the magnification parameter corresponding to each lens bin, and $C^{\mathrm{ij}}_{\mathrm{mm}}(\ell)$ is the convergence power spectrum between the lens and source distributions.  The last contribution to Equation \ref{eq_14} is the cross-correlation between lens magnification and source IA. This relation can be written as:
\begin{equation}\label{cross_mag_IA}
C^{\mathrm{ij}}_{\mathrm{mI}}(\ell) = \int d\chi \, \frac{q^i_l(\chi) N^j_s(\chi)}{\chi^2} \, P_{\mathrm{mI}} \left( k=\frac{\ell+1/2}{\chi},z(\chi)\right)
\end{equation}
where $P_{\mathrm{mI}} = P_{\mathrm{gI}}$.

\subsection{Parameterizations}
\label{sec_parameters}

In this section we summarise the parameterization that we adopt for modelling the $3 \times 2$-pt correlations and shear ratio. 

\begin{itemize}
    \item \textbf{Cosmological parameters}: we use 6 cosmological parameters to specify the $\Lambda$CDM model, which are the matter density $\Omega_\mathrm{m}$, Hubble parameter $H_0$, baryon density $\Omega_\mathrm{b}$, primordial spectral index $n_s$, neutrino density $\Omega_{\nu}$, and normalisation parameter $S_8$. We can derive $\sigma_8$ and hence the primordial power spectrum amplitude $A_s$ from $S_8$.  We assume three species of neutrinos and a flat universe.
    \item \textbf{Galaxy bias}: we assume a linear relation between the underlying dark matter density field and the lens galaxy density field. These lens galaxy bias parameters are denoted as $b_\mathrm{i}$, where $\mathrm{i}$ corresponds to the lens bin. Galaxy bias affects the $\gamma_\mathrm{t}(\theta)$ and $w(\theta)$ models, but not the shear ratio.  When we do parameter fitting we only vary the galaxy bias parameters when including the $3 \times 2$-pt correlations, and in this case we adopt wide, flat priors for these bias parameters.
    \item \textbf{Uncertainty of source redshift}: these parameters capture the uncertainty of distance measurements using photometric-redshift calibration methods. We denote these parameters as $\Delta z_\mathrm{j}$ where $\mathrm{j}$ corresponds to the source bin. We considered analyses including and excluding a prior in these source redshift distribution shifts.  For DES-Y3 and HSC-Y1, we used independent Gaussian priors specified by each survey. For DES, we used the values described by \citet*{2022PhRvD.105h3529S}, and for HSC, we adopted the priors of \cite{2022PASJ...74..923H}. For KiDS-1000, we used the correlated priors between the five KiDS tomographic bins \citep{2021A&A...647A.124H}. In this case we vary these uncorrelated parameters and derive the correlated $\Delta z$ shifts.
    \item \textbf{Intrinsic Alignment (IA) parameters}: For the NLA model \citep{2007Bridle} we use two free parameters: the tidal amplitude ($A_1$) and the power-law slope ($\eta_1$) in Equation \ref{eq_22}. For the TATT model \citep{2019Blazek} we use 5 parameters as described above: the tidal amplitude ($A_1$), the torquing amplitude ($A_2$), two parameters that relate to the power-law in Equations \ref{eq_22} and \ref{eq_23} ($\eta_1$ and $\eta_2$), and a bias parameter ($b_{TA}$) reflecting that galaxies are over-sampled in highly clustered regions.  We used flat priors for these parameters.
    \item \textbf{Multiplicative shear bias parameters}: these are the systematic parameters that determine the true ellipticity of sources from the observed shapes. We denote these parameters as $m_\mathrm{j}$, where $\mathrm{j}$ corresponds to source bins. We always used independent Gaussian priors for all surveys, which have been calibrated by image simulations. For DES-Y3 we adopted the specifications of \citet*{2022PhRvD.105h3529S} and \cite{2020MNRAS.491.5498M}, for HSC we used the priors described by \cite{2022PhRvD.106h3520M} and for KiDS, we applied the specification of \cite{2021A&A...645A.105G}.
    \item \textbf{Lens magnification parameters}: these parameters describe the sign and amplitude of the effect of lens magnification and are denoted as $\alpha_\mathrm{i}$, where $\mathrm{i}$ corresponds to the lens bin. We fix the magnification parameters based on the analysis of \cite{2021MNRAS.504.1452V}, where we used the low-redshift measurement for our lens bins 1, 2 and 3 and the high-redshift measurement for our lens bins 4 and 5. Hence we adopt magnification coefficients $\alpha = (1.93, 1.93, 1.93, 2.62, 2.62)$ for the five BOSS lens samples.
\end{itemize}

\subsection{Priors and fiducial values}

The errors that we obtain in parameter forecasts or fits are highly dependent on the priors adopted for these parameters.  The priors that we adopt in our analysis can be seen in Table \ref{table:priors}, and are similar to the choices presented by \citet*{2022PhRvD.105h3529S}. The Gaussian priors for $\Delta z$ and $m$ depend on the individual weak lensing survey as described above, and are listed in Table \ref{table:pri_surv}.  We use a different parameterization for KiDS because it has correlated $\Delta z$ priors, which are transformed to uncorrelated variables within the Bayesian inference \citep{2021A&A...647A.124H}.  The fiducial parameter values that we use for calculating the Fisher matrix can be seen in Table \ref{table:fiducial_value}.

\begin{table}
\centering
\begin{tabular}{c c c } 
 \hline
 Parameter & Range \\
 \hline
 \textbf{Systematic Parameters}\\
 Galaxy bias $b_\mathrm{i}$ & $[0.5, 5.0]$ \\ 
 $\Delta z$ & [-0.2, 0.2] \\
 IA $A_1$, $\eta_1$ (NLA) & $[-5.0, 5.0]$ \\
 IA $A_1$, $A_2$, $\eta_1$, $\eta_2$ (TATT) & $[-5.0, 5.0]$ \\
 IA $b_{\mathrm{TA}}$ (TATT) & $[0.0, 2.0]$ \\
 Shear calibration $m$ & [-0.1, 0.1] \\  
 \textbf{Cosmological Parameters}\\
 $\Omega_\mathrm{m}$ & $[0.1, 0.9]$ \\ 
 $S_8$ & $[0.1, 1.3]$ \\
 $h$ & $[0.55, 0.91]$ \\
 $\Omega_\mathrm{b}$ & $[0.03, 0.07]$ \\
 $n_s$ & $[0.87, 1.07]$ \\
 $\Omega_{\nu} h^2$ & $[0.0006, 0.00644]$ \\
 \hline
\end{tabular}
\caption{The set of systematic weak lensing and cosmological parameters we consider in our analysis, together with the uniform priors we adopt for each.  The parameters in this table apply to all weak lensing surveys.  We note that these wide, uniform priors in $\Delta z$ and $m$ are modified for each individual weak lensing survey by Gaussian priors, as described in Table \ref{table:pri_surv}.}
\label{table:priors}
\end{table}

\begin{table}
\centering
\begin{tabular}{c c } 
 \hline
 Parameter & Prior \\
 \hline
 \textbf{KiDS} \\
 uncorr $\Delta z_1$ & $\mathcal{N}(0.000, 1.0)$ \\
 uncorr $\Delta z_2$ & $\mathcal{N}(-0.181, 1.0)$ \\
 uncorr $\Delta z_3$ & $\mathcal{N}(-1.110, 1.0)$ \\
 uncorr $\Delta z_4$ & $\mathcal{N}(-1.395, 1.0)$ \\
 uncorr $\Delta z_5$ & $\mathcal{N}(1.265, 1.0)$ \\
 $m_1$ & $\mathcal{N}(-0.009, 0.019)$ \\ 
 $m_2$ & $\mathcal{N}(-0.011, 0.020)$ \\
 $m_3$ & $\mathcal{N}(-0.015, 0.017)$ \\
 $m_4$ & $\mathcal{N}(0.002, 0.012)$ \\
 $m_5$ & $\mathcal{N}(0.007, 0.010)$ \\
 \textbf{DES} \\
 $\Delta z$ & $\mathcal{N}(0,[0.018,0.015,0.011,0.017])$\\
 $m_1$ & $\mathcal{N}(-0.0063, 0.0091)$ \\ 
 $m_2$ & $\mathcal{N}(-0.0198, 0.0078)$ \\
 $m_3$ & $\mathcal{N}(-0.0241, 0.0076)$ \\
 $m_4$ & $\mathcal{N}(-0.0369, 0.0076)$ \\
 \textbf{HSC} \\
 $\Delta z$ & $\mathcal{N}(0 ,[0.0374, 0.0124, 0.0326, 0.0343])$\\
 $m$ & $\mathcal{N}(0,[0.01, 0.01, 0.01,0.01])$\\
 \hline
\end{tabular}
\caption{Priors specific to each weak lensing survey. For KiDS, we adopt a special parameterization in terms of uncorrelated $\Delta z$ parameters. }
\label{table:pri_surv}
\end{table}

\begin{table}
\centering
\begin{tabular}{cc}
\hline
Parameter & Fiducial value  \\
\hline
$\Delta z_\mathrm{i}$ & 0.0 \\
$A_1$, $A_2$ & 0.0, 0.0 \\
$m_\mathrm{i}$ & 0.0 \\
 $\Omega_\mathrm{m}$ & 0.286\\ 
 $S_8$ & 0.8128\\
 $h$ & 0.7\\
 $\Omega_\mathrm{b}$ & 0.046\\
 $n_s$ & 0.96\\
 $\Omega_{\nu} h^2$ & 0.00083\\
\hline
\end{tabular}
\caption{The fiducial values of the model parameters adopted in our Fisher matrix forecast analysis.}
\label{table:fiducial_value}
\end{table}

\section{Fisher Matrix Forecast}
\label{sec:fisher}

\subsection{Methodology}

Fisher matrix formalism is a standard tool for forecasting the statistical ability of experiments to measure parameters.  Suppose we have a data vector of length $N_\mathrm{data}$, which has covariance $\mathbf{C}$ with size $N_\mathrm{data} \times N_\mathrm{data}$. The data are fit by a model vector $\mathbf{m}$, which has the same length $N_\mathrm{data}$, and this model is a function of $N_\mathrm{par}$ parameters. The elements of the Fisher matrix $\mathbf{F}$ of the parameters, of size $N_\mathrm{par} \times N_\mathrm{par}$, can be written as:
\begin{equation}\label{Fisher}
F_{\mathrm{i,j}} = \frac{\partial \mathbf{m}^T}{\partial p_\mathrm{i}} \mathbf{C}^{-1} \frac{\partial \mathbf{m}}{\partial p_\mathrm{j}}
\end{equation}
In the limit of Gaussian likelihood, the parameter covariance matrix is the inverse of the Fisher matrix, and the expected error on the parameter $p_\mathrm{i}$ can be calculated by $(\mathbf{F}^{-1}_{\mathrm{i,i}})^{1/2}$. The Cramer-Rao bound defines that no unbiased estimator of the parameter performs better than the estimated error from the Fisher matrix; therefore, the Fisher information provides a lower limit on the error.

We use the Fisher matrix to forecast the error of weak lensing systematic parameters and cosmological parameters, including and excluding shear ratio data, allowing us to systematically consider the degree to which the shear ratio can improve constraints. After we compute the parameter covariances for each case, we determine the improvement in the forecast errors due to SR. We calculate the Fisher matrix for all weak lensing surveys combined with BOSS data.

The two-point correlation data vectors that we consider for this work are  $1 \times 2$-pt (that is, just cosmic shear) and $3 \times 2$-pt (that is, $\left[ \xi_+(\theta), \xi_-(\theta), \gamma_\mathrm{t}(\theta), w(\theta) \right]$), which we combined with SR using a fitting range $2 - 6 \, h^{-1}$ Mpc as described in Section \ref{sec:shearratio}.  We computed the analytical covariances of these sets of statistics using the methods described by \cite{Blake_C} and \cite{Blake_DESI}, as outlined in Section \ref{sec:gammat} and evaluated all models using the {\small COSMOSIS} platform adopting the source and lens redshift distributions introduced in Section \ref{sec:data}, and the parameterisations and assumptions described in Section \ref{sec:shearmodel}.

We consider two cases for Fisher matrix analysis: fixed and varying cosmology.  When studying fixed cosmology, we want to ascertain if SR can help measure the systematic astrophysical parameters of weak lensing, such as redshift distribution shifts and intrinsic alignment amplitudes.  Therefore, we also investigate cases with and without including the prior on $\Delta z$ (to test whether SR can serve as a useful alternative to this prior).  When varying cosmology, we want to understand the benefit of SR when all parameters are jointly free to vary.

We determine $\partial \mathbf{m}/\partial p$ numerically by evaluating the model at parameter values slightly offset from the fiducial values listed in Table \ref{table:fiducial_value} by $\pm 0.01$, except for the $\Omega_{\nu} h^2$ parameter, for which we use offsets of $\pm 0.0001$.  For the fixed cosmology forecast we use the prior in the shear calibration $m$, and for varying cosmology we also adopt Gaussian priors in four of the cosmological parameters with standard deviations $[\sigma_h, \sigma_{\Omega_\mathrm{b}}, \sigma_{n_s}, \sigma_{\Omega_{\nu} h^2}] = [0.09,0.01,0.05,0.00146]$.  These standard deviations represent half the width of the uniform priors used in the MCMC analysis presented in Table \ref{table:priors}, except they are expressed as Gaussian priors.  These can be readily combined in a Fisher matrix forecast by adding the prior Fisher matrix to the existing parameter Fisher matrix as in Equation \ref{Fis_prior},
\begin{equation}\label{Fis_prior}
F_{\mathrm{final}} = F_{\mathrm{i,j}} + F_{\mathrm{prior}}
\end{equation}
where $F_{\mathrm{prior}}$ is expressed as,
\begin{equation}
F_{\mathrm{prior}} =
    \begin{bmatrix}
     1/\sigma_1^2 & 0 & 0 & \cdots & 0 \\
     0 & 1/\sigma_2^2 & 0 & \cdots & 0 \\
     0 & 0 & 1/\sigma_3^2 & \cdots & 0 \\
     \vdots & \vdots & \vdots & \ddots & \vdots \\
     0 & 0 & 0 & \cdots & 1/\sigma^2_{\mathrm{N}}\\
     \end{bmatrix}
\end{equation}
and $\sigma_1, \sigma_2, \sigma_3, \cdots, \sigma_{\mathrm{N}}$ are the standard deviations of the priors in each parameter.

\subsection{Results}

The results of our Fisher matrix forecasts are displayed in Table \ref{table:Fisher Matrix}, quantifying the improvement in measuring different parameters across the cases we consider.  For our analysis at fixed cosmology, we forecast that including (or excluding) the prior in the source redshift distribution shifts $\Delta z$, motivated by photometric redshift calibration techniques, has a significant impact on the relative contribution of SR data to cosmic shear data.  Since this prior is something we may wish to independently validate, we repeat our forecast with and without the $\Delta z$ prior.

\begin{table*}
\centering
\begin{tabular}{ccccc}
\hline
WL survey & Parameter & Cosmology & Improvement & Improvement \\
& & & without $\Delta z$ prior & with $\Delta z$ prior  \\
\hline

KiDS-1000 & $\Delta z_\mathrm{i}$ &  Fixed & 9.8 - 37.7\% &   0.1 - 1.9\% \\
DES-Y3   &  $\Delta z_\mathrm{i}$ &    Fixed & 13.6 - 30.3\% &   1.6 - 4.6\% \\
HSC-Y1   & $\Delta z_\mathrm{i}$ &    Fixed  & 3.8 - 40.6\% &    2.0 - 15.8 \% \\
                
KiDS-1000 & $A_1$   & Fixed &   43.8\% &   14.4\% \\
DES-Y3 & $A_1,A_2$ &    Fixed   & 32.1, 13.8\% &    9.5, 4.2\% \\
HSC-Y1 & $A_1,A_2$ &    Fixed   & 34.7, 13.2\%  & 23.8, 12.0\% \\

\hline

KiDS-1000 & $\Delta z_\mathrm{i}$ &  Varying &   32.8 - 40.9\% & 0.1 - 2.3\% \\
DES-Y3   &  $\Delta z_\mathrm{i}$ &    Varying & 14.8 - 35.7\% & 1.0 - 6.9\% \\
HSC-Y1   & $\Delta z_\mathrm{i}$ &    Varying    & 53.2 - 70.1\% &   1.7 - 15.6 \% \\
                
KiDS-1000 & $A_1$   & Varying & 45.0\% &  21.3\% \\
DES-Y3 & $A_1,A_2$ &    Varying & 24.4, 6.0\% & 13.3, 4.0\% \\
HSC-Y1 & $A_1,A_2$ &    Varying & 34.2, 17.3\%  & 32.0, 16.6\% \\

\hline

KiDS-1000 & $\Omega_\mathrm{m}$ &    Varying &   23.0\% &    6.2\% \\
DES-Y3   & $\Omega_\mathrm{m}$ & Varying &10.4\% &   3.5\% \\
HSC-Y1 & $\Omega_\mathrm{m}$ &   Varying & 9.3\% &   12.4\% \\
                
KiDS-1000 & $S_8$ & Varying & 35.5\%    & 0.2\% \\
DES-Y3 & $S_8$  & Varying   & 39.9\%    & 2.7\% \\
HSC-Y1 & $S_8$  & Varying   & 60.2\%    & 2.1\% \\

\hline
\end{tabular}
\caption{The improvement in the determination of selected model parameters forecast by the Fisher matrix analysis, following the addition of shear ratio (SR) data to cosmic shear ($1 \times 2$-pt) data. Inclusion of the prior in $\Delta z$ (from photometric redshift calibration) is also significant in determining the relative contribution of shear ratio data.  We indicate if the cosmological parameters are held fixed or varying in each forecast.}
\label{table:Fisher Matrix}
\end{table*}

\begin{figure*}
    \centering
    \includegraphics[width=0.9\linewidth]{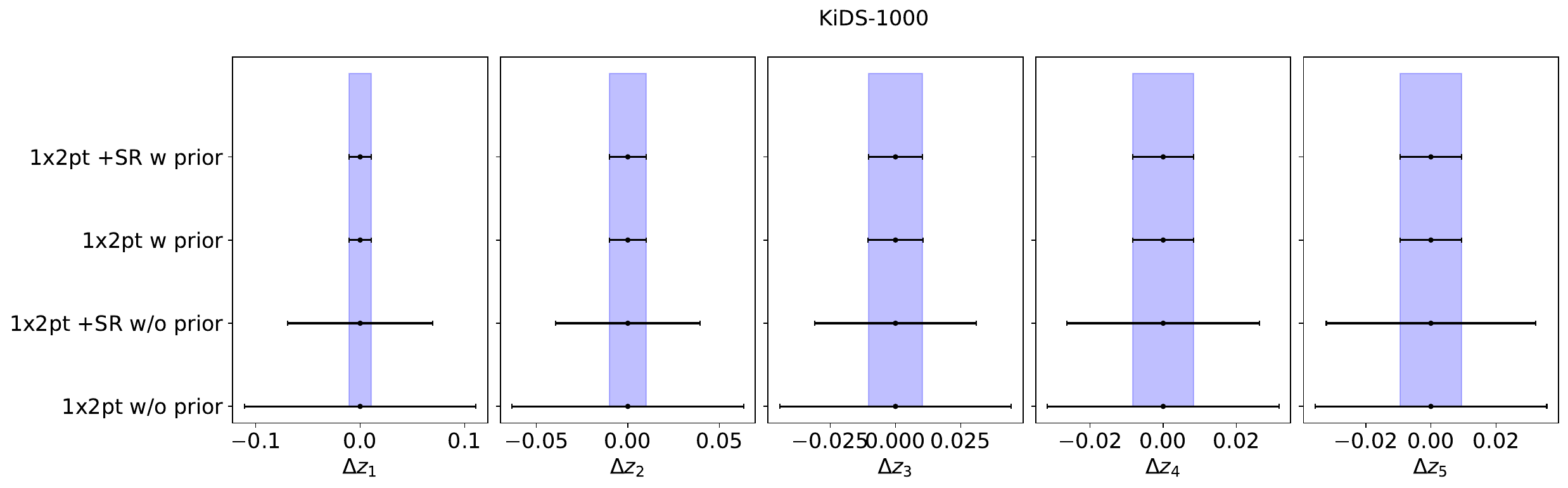}
    \vskip\baselineskip
    \includegraphics[width=0.8\linewidth]{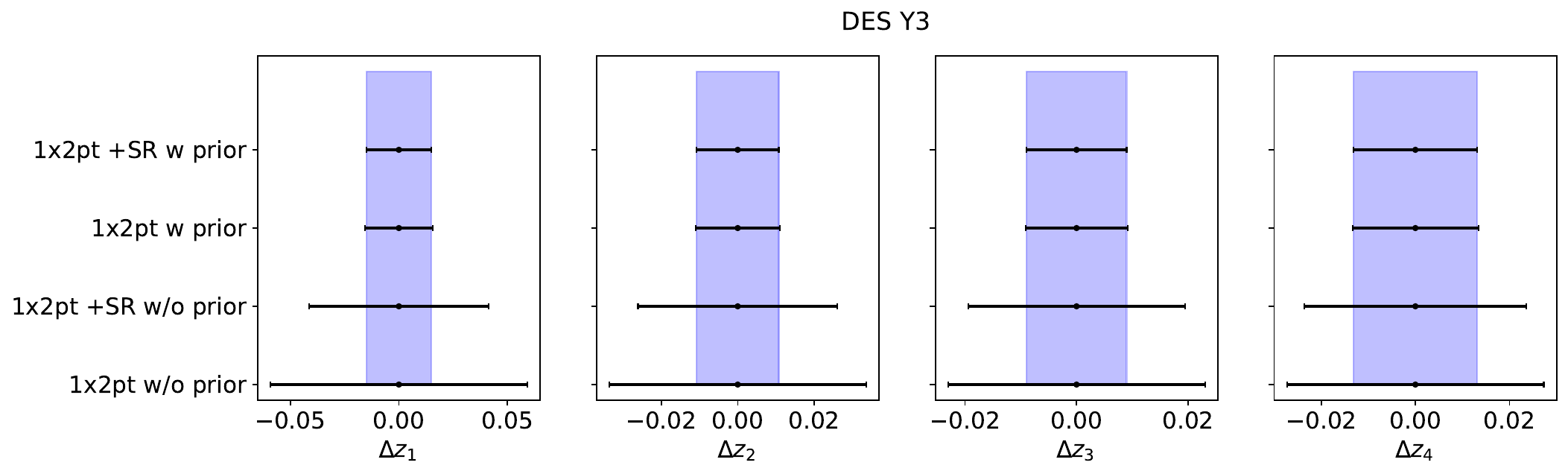}
    \vskip\baselineskip
    \includegraphics[width=0.8\linewidth]{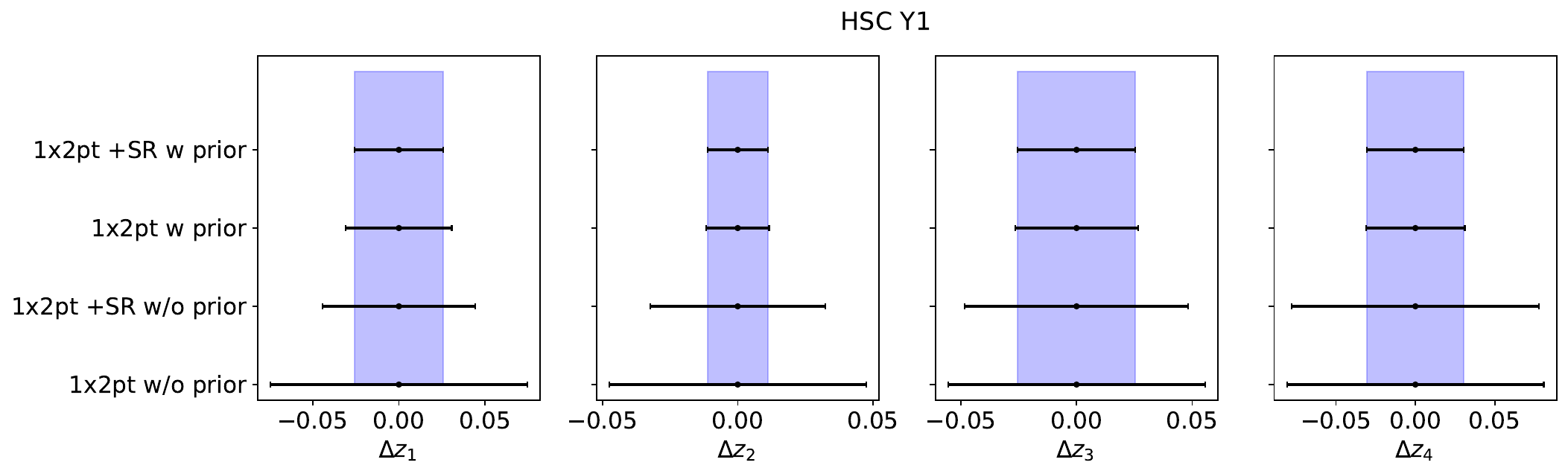}
    \caption{Forecast errors in $\Delta z$ from the Fisher matrix analysis for fixed cosmology, with and without the prior in $\Delta z$. Without the $\Delta z$ prior the forecast errors in these parameters are large, but show significant improvement when $1 \times 2$-pt data are combined with SR. Including the $\Delta z$ prior reduces the error and applying SR in this case still improves the forecast constraint in $\Delta z$, but to a smaller extent.}
    \label{comp_FM}
\end{figure*}

We focus first on forecasts obtained for constraining the redshift distribution shifts $\Delta z$ by combining SR with $1 \times 2$-pt data, which can be seen in Figure \ref{comp_FM}.  Excluding the $\Delta z$ prior, we obtain forecast errors from cosmic shear for $\Delta z$ in the tomographic bins of $[0.03 - 0.11]$ for KiDS-1000, $[0.02 - 0.06]$ for DES-Y3 and $[0.05 - 0.08]$ for HSC-Y1 (where we quote the minimum and maximum errors across the different tomographic bins). After combining $1 \times 2$-pt with SR, we obtain a significant improvement in these forecasts of up to 38\% for KiDS-1000, up to 30\% for DES-Y3 and up to 40\% for HSC-Y1.

After applying the $\Delta z$ prior, the forecast error in $\Delta z$ significantly improves, and the additional information from SR diminishes.  Including the prior, the $\Delta z$ error ranges are $[0.008 - 0.011]$ for KiDS-1000, $[0.009 - 0.016]$ for DES-Y3, and $[0.012 - 0.031]$ for HSC-Y1.  After combining with SR, we obtain an improvement of up to 2\% for KiDS-1000, up to 5\% for DES-Y3 and up to 16\% for HSC-Y1. We conclude from this analysis that whilst photometric-redshift calibration information is more accurate than SR, shear ratio data still contains significant information about $\Delta z$ that can improve the constraints.  Generally, we find that the lowest-redshift source bins contain greatest improvement.

Next, we consider the forecast constraints on the IA parameters within the NLA model for KiDS-1000 and the TATT model for DES-Y3 and HSC-Y1, where we focus on the amplitude parameters as the metric for the IA constraints, and again consider results with and without the $\Delta z$ prior.  Without the $\Delta z$ prior, we forecast an error in the IA parameters from cosmic shear only of $\sigma[A_1] = [0.23]$ for KiDS-1000, $\sigma[A_1, A_2] = [0.60, 0.92]$ for DES-Y3 and $\sigma[A_1, A_2] = [1.18, 2.35]$ for HSC-Y1. After applying SR,  this constraint improves by 44\% for the $A_1$ parameter for KiDS-1000, and the improvements for DES-Y3 are $[32, 14]\%$, and $[35, 13]\%$ for HSC-Y1, for $[A_1, A_2]$ respectively.

\begin{figure}
    \centering
    \includegraphics[width=0.8\linewidth]{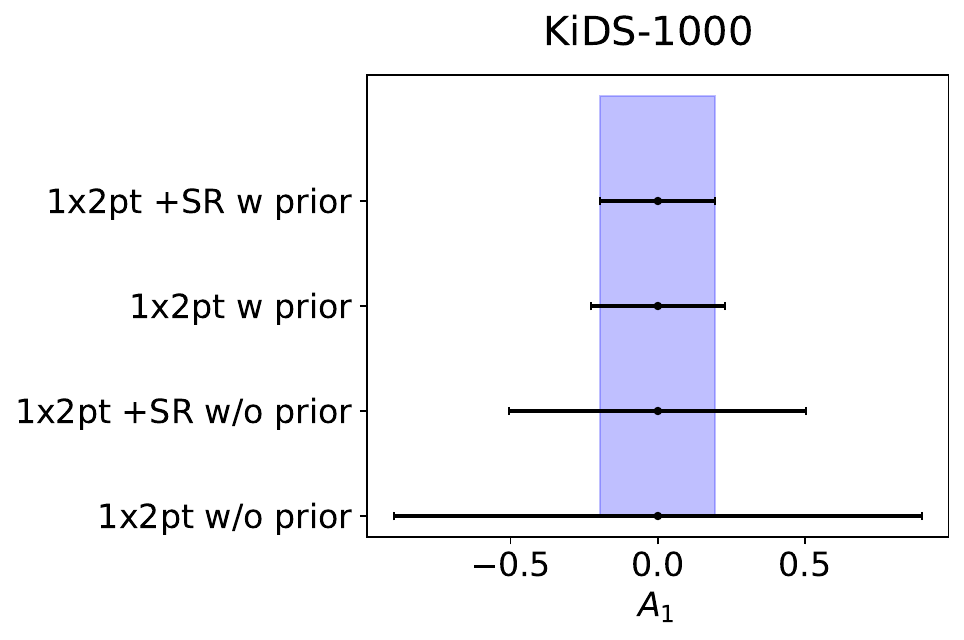}
    \vskip\baselineskip
    \includegraphics[width=0.8\linewidth]{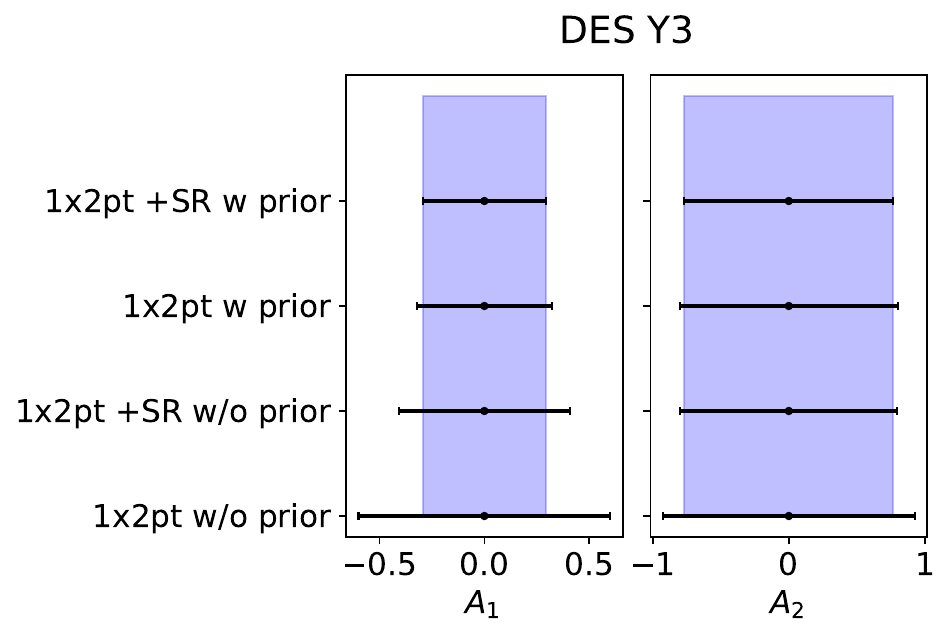}
    \vskip\baselineskip
    \includegraphics[width=0.8\linewidth]{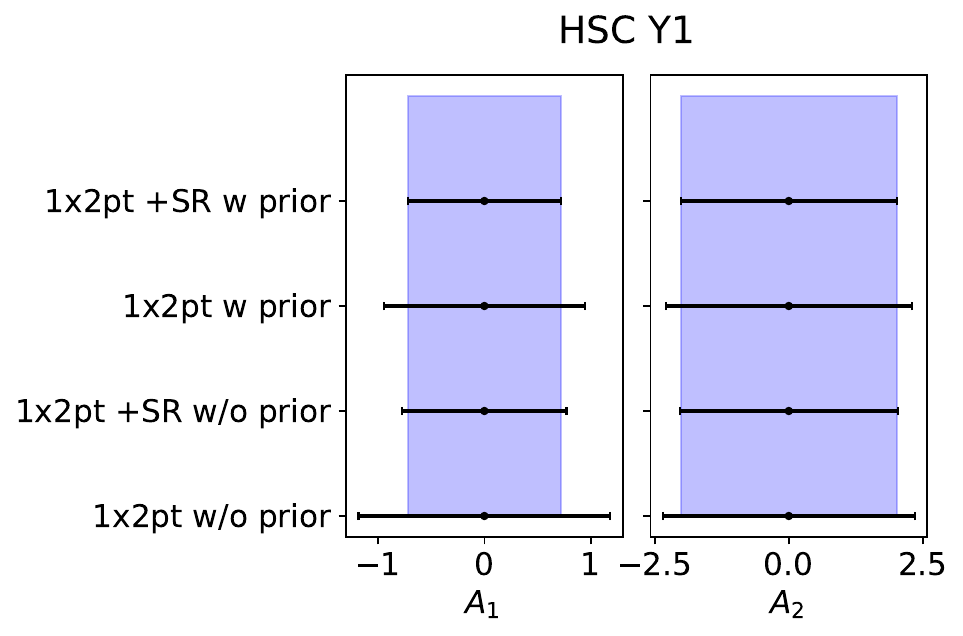}
    \caption{Forecast errors in the IA amplitudes $A_1$ and $A_2$ from the Fisher matrix analysis for fixed cosmology, with and without the prior in $\Delta z$.  In this case the shear ratio can play a comparable role in improving parameter constraints as the $\Delta z$ prior, when combined with cosmic-shear only data.}
    \label{comp_FM_2}
\end{figure}

When we include the $\Delta z$ prior for KiDS-1000, we forecast an additional improvement of up to 75\% in determining the IA amplitude.  For DES-Y3 and HSC-Y1, including the $\Delta z$ prior makes the forecast errors in IA parameters from cosmic shear improve by up to 46\% (compared to without the prior).  Interestingly, for DES-Y3 and HSC-Y1, the inclusion of SR has more effect on the determination of $A_2$ than adding a $\Delta z$ prior.  The comparison between these forecast errors for IA parameters can be seen in Figure \ref{comp_FM_2}.  Adding SR to the $\Delta z$ prior, the forecast error in the IA parameters improves by 14\% for KiDS for $A_1$, and $[9, 4]\%$ for DES-Y3 and $[24, 12]\%$ for HSC-Y1, for $[A_1, A_2]$ respectively.

We now consider forecasts varying both astrophysical and cosmological parameters, in which we also consider analyses with and without the $\Delta z$ prior. The forecast errors in the $\Delta z$ and IA parameters are comparable (albeit slightly larger) to those obtained for fixed cosmology, but they show differing improvements after we include SR.  Without the $\Delta z$ prior, the forecast improvement in $\Delta z$ after including SR ranges up to 41\% for KiDS-1000, 36\% for DES-Y3 and 70\% for HSC-Y1.  With the $\Delta z$ prior, these improvements are up to 2\% for KiDS-1000, 7\% for DES-Y3 and 16\% for HSC-Y1.

For the IA parameter for KiDS-1000, when we analyse without the $\Delta z$ prior we forecast a 45\% improvement from adding SR. For DES-Y3 and HSC-Y1 without the $\Delta z$ prior, the forecast improvements in the errors in $[A_1, A_2]$ after applying SR are  $[24, 6]\%$ for DES-Y3, and $[34, 17]\%$ for HSC-Y1. With the $\Delta z$ prior, the improvements are $21\%$ for KiDS-1000, $[13, 4]\%$ for DES-Y3, and $[32, 17]\%$ for HSC-Y1.  In summary, the importance of the shear ratio to the forecast constraints increases once more parameters are varying.

\begin{figure}
    \includegraphics[width=0.8\linewidth]{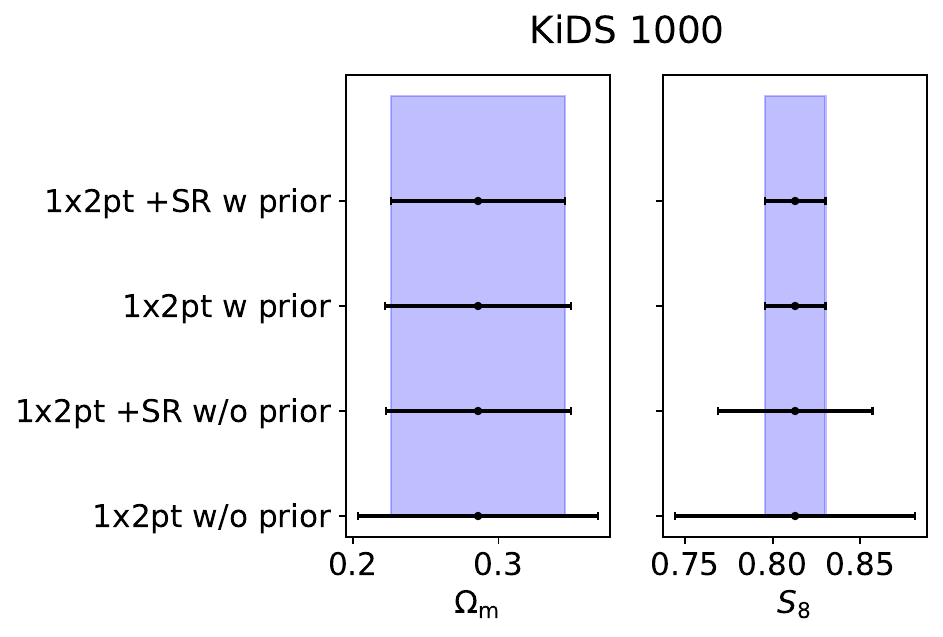}
    \vskip\baselineskip
    \includegraphics[width=0.8\linewidth]{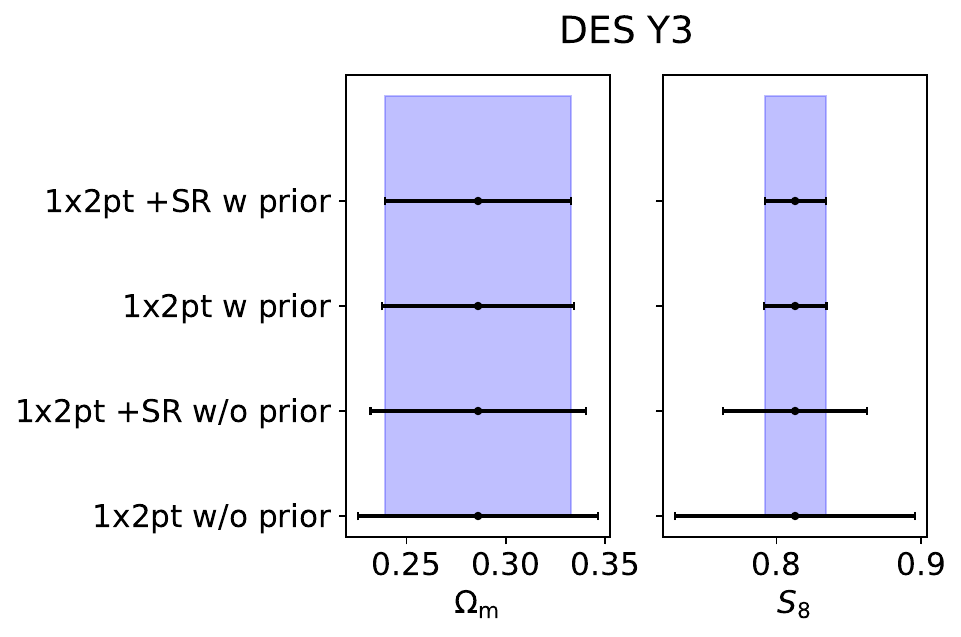}
    \vskip\baselineskip
    \includegraphics[width=0.8\linewidth]{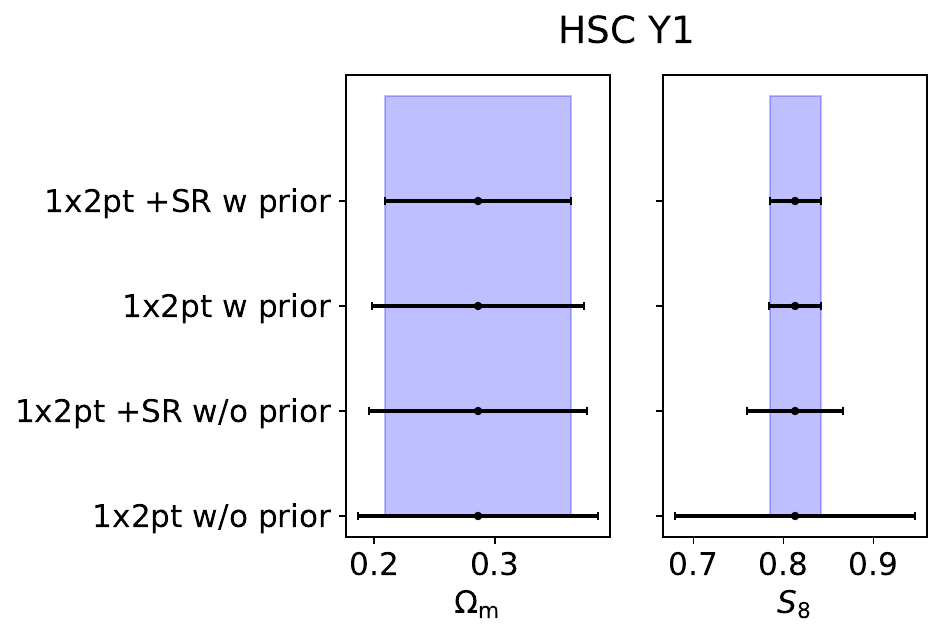}
    \caption{Forecast errors in the cosmological parameters $\Omega_\mathrm{m}$ and $S_8$ from the Fisher matrix analysis for varying cosmology.  We also vary $(\Omega_\mathrm{b}, h, n_s, \Omega_\nu h^2)$, with and without the prior in $\Delta z$.  We find that the $\Delta z$ prior has a significant effect on the forecast errors, especially for $S_8$.  Including SR data improves the constraint on cosmological parameters when the $\Delta z$ prior is unavailable, compared to cosmic-shear only data.}
    \label{comp_FM_3}
\end{figure}

We also investigated the forecast improvement when including SR of constraints on two key cosmological parameters, $\Omega_\mathrm{m}$ and $S_8$.  These results are displayed in Figure \ref{comp_FM_3}.  First we consider the case without the $\Delta z$ prior. The forecast errors in $[\Omega_\mathrm{m}, S_8]$ are $[0.08, 0.07]$ for KiDS-1000, $[0.06, 0.08]$ for DES-Y3, and $[0.10, 0.13]$ for HSC-Y1. After applying SR, the improvement in these error forecasts are $[23, 36]\%$ for KiDS-1000, $[10, 40]\%$ for DES-Y3, and $[9, 60]\%$ for HSC-Y1.

The forecast errors in $\Omega_\mathrm{m}$ and $S_8$ become narrower once we include the $\Delta z$ prior: $\sigma[\Omega_\mathrm{m}, S_8]$ are $[0.06, 0.02]$ for KiDS-1000, $[0.05, 0.02]$ for DES-Y3, and $[0.09, 0.03]$ for HSC-Y1.  The improvement after applying SR is dramatically smaller than without the $\Delta z$ prior: $[6.2, 0.2]\%$ for KiDS-1000, $[4, 3]\%$ for DES-Y3, and $[12, 2]\%$ for HSC-Y1, for $[\Omega_\mathrm{m}, S_8]$ respectively.  When both astrophysical and cosmological parameters are varying, we conclude that including or excluding the $\Delta z$ prior has an important impact on the relative improvement when combining SR with cosmic shear data. 

We end by forecasting the comparative performance when $3 \times 2$-pt statistics are used as the dataset, rather than just cosmic shear.  In this test we always apply the $\Delta z$ prior and vary all parameters, including cosmology.  As expected, the improvement when adding shear ratio information is significantly reduced when considering a baseline of $3 \times 2$-pt statistics, although it can still marginally help constrain these parameters. Applying SR produces small improvements in $\Delta z$ of up to $[1.8, 5.2, 10]\%$ for KiDS-1000, DES and HSC, respectively. For IA parameters, after we add the SR to KiDS-1000 $3 \times 2$-pt data, we find an improvement in $A_1$ of $10.7\%$.  We also obtain the forecast improvements for DES-Y3 and HSC-Y1 of $[11.1, 4.3]\%$  and $[20.5, 10.7]\%$, for $[A_1, A_2]$ respectively. Finally, the forecast errors in $\Omega_\mathrm{m}$ and $S_8$ are tighter in the $3 \times 2$-pt analysis than for the $1 \times 2$-pt correlations. After we apply the SR, we obtain small improvements in $[\Omega_\mathrm{m}, S_8]$ of $[0.8, 0.5]\%$ for KiDS-1000, $[0.6, 2.6\%]$ for DES-Y3 and $[1.3, 0.4]\%$ for HSC-Y1. We conclude from this analysis that for the configurations we are considering, shear ratio information does not significantly contribute to cosmological parameter measurements when reliable $3 \times 2$-pt information and the $\Delta z$ prior is available.

\subsection{Variation of SR contribution with survey properties}

We used our Fisher forecast platform to investigate the contribution of the shear ratio to cosmological and astrophysical parameter determinations as we varied three survey properties: the minimum scale included in the SR measurements, the lens number density, and the prior in the mean source redshift distribution.  Hence, we aim to quantify the conditions and parameters for which the SR contributes most information to $1 \times 2$-pt and $3 \times 2$-pt correlations.  We otherwise used the same BOSS, KiDS-1000, DES-Y3 and HSC-Y1 configurations described in the previous sections.

First, we checked the impact of including SR after extending the scale range we utilise from $2-6 \, h^{-1}$ Mpc to $0.5-6 \, h^{-1}$ Mpc (noting that more detailed tests with simulations would be required to establish the validity of models at these smaller scales).  We find that the level of improvement depends significantly on the source density of the lensing survey, which directly influences the small-scale error, hence we find the largest benefits for the HSC survey data.  The most significant improvements we find in this case are for the IA amplitudes $[A_1, A_2]$, where the addition of shear ratio now results in improvements relative to the baseline of 42.8\% - 66.4\% (for $1 \times 2$-pt) and 35.3\% - 56.8\% (for $3 \times 2$-pt).  However, the improvements in the determination of cosmological parameters are less significant, at the 1\% level for $S_8$.  We conclude that fitting shear ratio data to smaller scales can significantly improve our determination of IA parameters, but not directly cosmological parameters.



Second, we considered a model lens survey with $10\times$ the galaxy number density of BOSS, returning to the fiducial fitting range $2-6 \, h^{-1}$ Mpc, and re-generating the analytical covariance matrix with the increased lens density.  This configuration again improves the galaxy-galaxy lensing error at small scales.  Once more, we find the most significant benefits from adding SR data are in the error in determining IA parameters, for which we find improvements in the range 9.7\% - 49.1\%, depending on the lensing survey and baseline statistic.  Once more, improvements in cosmological parameter determinations were less significant.



\begin{figure}
  \includegraphics[width=\linewidth]{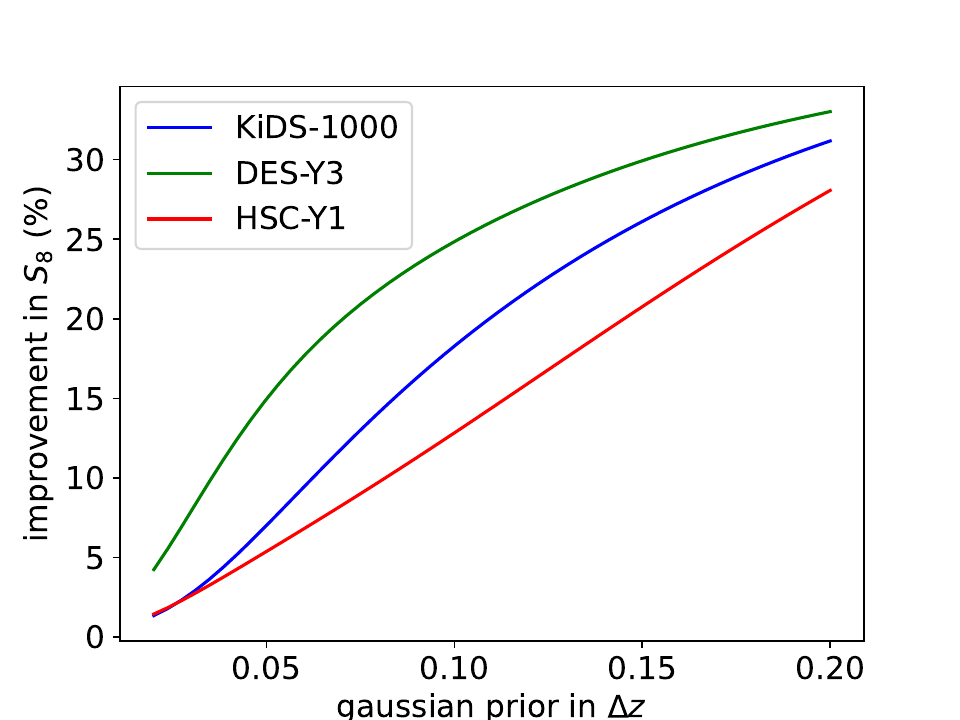}
  \caption{The forecast improvement in the error in determining the $S_8$ parameter after adding SR measurements to cosmic shear data, as a function of the width of the Gaussian prior adopted for the mean source redshift $\Delta z$.  Forecasts are shown for the KiDS-1000, DES-Y3 and HSC-Y1 lensing surveys, in combination with BOSS.  We find that the improvement resulting from SR information is highly dependent on our knowledge of the source redshift distribution.}
  \label{fig:Dz_priorallsurv}
\end{figure}

Finally, we investigated the improvement in the determination of $S_8$ when adding shear ratio information to cosmic shear ($1 \times 2$pt) data, as a function of the width of the Gaussian prior $\Delta z$, which we varied in the range $0.02 - 0.2$, with all other settings restored to fiducial.  The results are displayed for KiDS-1000, DES-Y3 and HSC-Y1 in Figure \ref{fig:Dz_priorallsurv}.  We can see that the improvement in $S_8$ is highly dependent on the availability and reliability of the $\Delta z$ prior.  In its absence, SR contributes significant new information to cosmological parameters.

\section{Bayesian Inference}
\label{sec:mcmc}

\subsection{Methodology}

Having established the forecasts, we now perform a Bayesian likelihood analysis to fit the model parameters to our datasets.  Starting from the average tangential shear and shear ratio measurements presented in Section \ref{sec:measurements}, we add the shear correlation function datasets for $[\xi_+(\theta), \xi_-(\theta)]$ publicly-released by the KiDS-1000, DES-Y3 and HSC-Y1 weak lensing collaborations, and (in some cases) the $w(\theta)$ angular correlation function measurements for BOSS presented by \cite{Blake_C}.  As described in Section \ref{sec:gammat}, we generated an analytical covariance matrix spanning these correlation function observables using the same methods as outlined in \cite{Blake_C} and \cite{Blake_DESI}.

Mathematically, Bayes' inference equation can be written as:
\begin{equation}
\label{Bayesian}
P(p|\mathbf{d}) \propto \mathcal{L}(\mathbf{d}|p) \, \Pi(p)
\end{equation}
where $p$ is the parameter space, $\mathbf{d}$ is the data, $P(p|\mathbf{d})$ is the posterior probability, $\mathcal{L}(\mathbf{d}|p)$ is the likelihood, and $\Pi(p)$ is the prior probability. We adopt a Gaussian likelihood written as:
\begin{equation}\label{Likelihood}
\mathcal{L}(\mathbf{d}|p) \propto -\frac{1}{2} \left[ \mathbf{d}-\mathbf{m}(p) \right]^T \mathbf{C}^{-1} \left[ \mathbf{d}-\mathbf{m}(p) \right]
\end{equation}
where $\mathbf{m}(p)$ is the model and $\mathbf{C}$ is the data covariance matrix. 

We implement Bayesian inference using nested sampling, which is a widely-used algorithm to sample the posterior probability of high-dimensional parameter spaces.  Nested sampling maps the high-dimensional posterior into one-dimensional space and divides the posterior into many nested slices.  In our case, we use {\small MULTINEST} \citep{Multinest1, Multinest2, Multinest3} within the {\small COSMOSIS} framework \citep{2015A&C....12...45Z}.  We use {\small MULTINEST} settings $n_\mathrm{live} = 1000$, $n_\mathrm{eff} = 0.3$ and $\mathrm{tol} = 0.01$.

We apply scale cuts in our fitting.  For the cosmic shear data, we use the same scale cuts as adopted by each weak lensing survey collaboration. For KiDS-1000 we use $[0.5 - 300']$ for $\xi_+$ \citep{kids1000_cut1}, which suppresses small-scale information when the contribution from baryon feedback is greater than $20\%$ of the overall signal. For $\xi_-$ we use scales $\theta >  4'$, excluding small physical scales where the modelling becomes challenging \citep{kids1000_cut2}. For DES-Y3 we use scale cuts based on a $\Delta \chi^2$ analysis using noiseless synthetic cosmic shear data vectors, generated with and without a baryonic contamination (see \cite*{DES_cut} for details), which leads to a different scale cut for every tomographic sample based on this threshold.  HSC-Y1 uses scales in the range $[7.08 - 56.2']$ and $[28.2 - 178']$ for $\xi_+$ and $\xi_-$, respectively, based on baryonic physics, the extra shape correlation due to PSF leakage, and PSF model error and signal-to-noise \citep{hsc_cut}.  When we also use $\gamma_\mathrm{t}(\theta)$ and $w(\theta)$ in the analysis, we fit to projected separations $R > 6 \, h^{-1}$ Mpc (converted to angular scales at each lens redshift slice).  For shear ratios we use a fitting range $2 - 6 \, h^{-1}$ Mpc, which does not overlap with the scales used for $\gamma_\mathrm{t}$, as discussed in Section \ref{sec:shearratio}.  Shear ratios have the potential to fit to smaller scales because non-linear effects may cancel out \citep*{2022PhRvD.105h3529S}.

\subsection{Results}

We used {\small COSMOSIS} to perform MCMC fits for weak lensing systematic parameters and cosmological parameters, following similar cases to those we forecast in Section \ref{sec:fisher}.  Our results are summarised in Table \ref{error_mc} and the $\chi^2$ values of the best fits are listed in Table \ref{chi_mcmc}.  We note that we obtain similar $\chi^2$ values for our KiDS-1000 and DES-Y3 analyses as previous works \citep*{kids1000_cut2, DES_cut}.  When interpreting the comparison between the Fisher matrix forecasts and MCMC results, we note that the Fisher matrix analysis assumes Gaussian joint confidence intervals between parameters, an approximation which may become inaccurate for several cases of interest, such that the MCMC analysis provides more realistic confidence regions.  On the other hand, the MCMC fits contain noise within each realisation, which will ``scatter'' the resulting improvements with respect to any Fisher matrix forecast of the ``average'' improvement.

Focussing first on the $1 \times 2$-pt fits, we again consider cases of fixed and varying cosmology, excluding and including the $\Delta z$ prior.  Applying this prior improves all parameter measurements, as we forecast previously.  In some cases, such as the $\Delta z$ parameters in KiDS-1000, the MCMC analysis produces smaller parameter errors than forecast by the Fisher matrix, which we attribute to the fact that the priors are not fully equivalent between the two analyses.  Considering the benefit of adding shear ratio data to the cosmic shear, we find improvements for some parameters, although not all.  We display the MCMC results for the improvements in Table \ref{table:MC_result_improve}, which mirrors the format of the forecasts we presented in Table \ref{table:Fisher Matrix}.  We briefly summarise these results as follows.

Considering the $\Delta z$ parameter fits for fixed cosmology and without a $\Delta z$ prior: the constraints for DES-Y3 and HSC-Y1 show significant improvements when adding SR, up to $27\%$ for DES-Y3 and up to $24\%$ for HSC-Y1. For the KiDS-1000 results, some $\Delta z$ measurements improve by up to $12\%$, and some do not.  As expected, the improvements when adding SR data are significantly lower for $\Delta z$ when the prior is available.  For the IA parameters for fixed cosmology, the benefit of adding SR information to cosmic shear is mixed.

%

\begin{figure}
    \centering
    \includegraphics[width=0.9\linewidth]{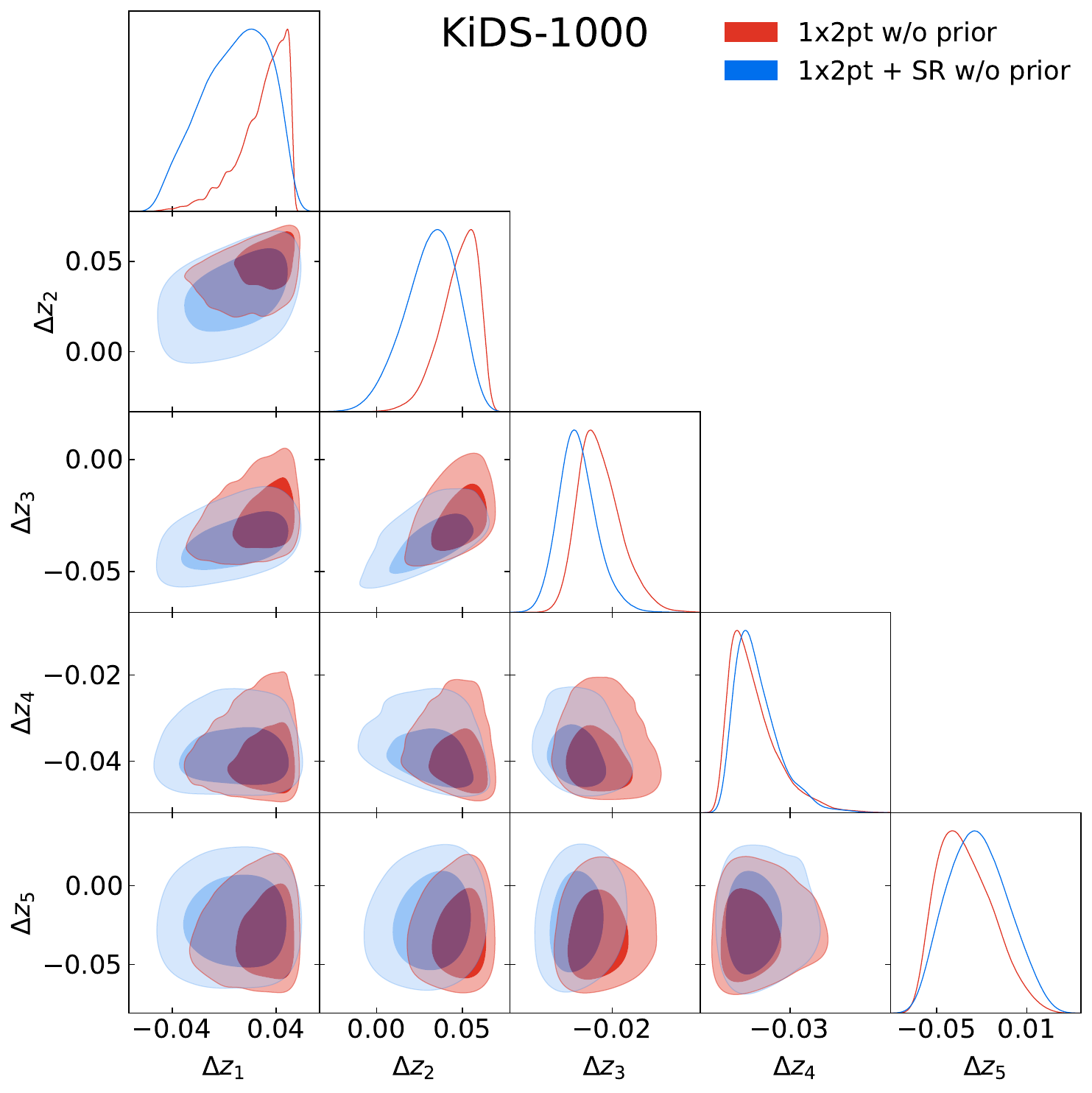}
    \vskip\baselineskip
    \includegraphics[width=0.9\linewidth]{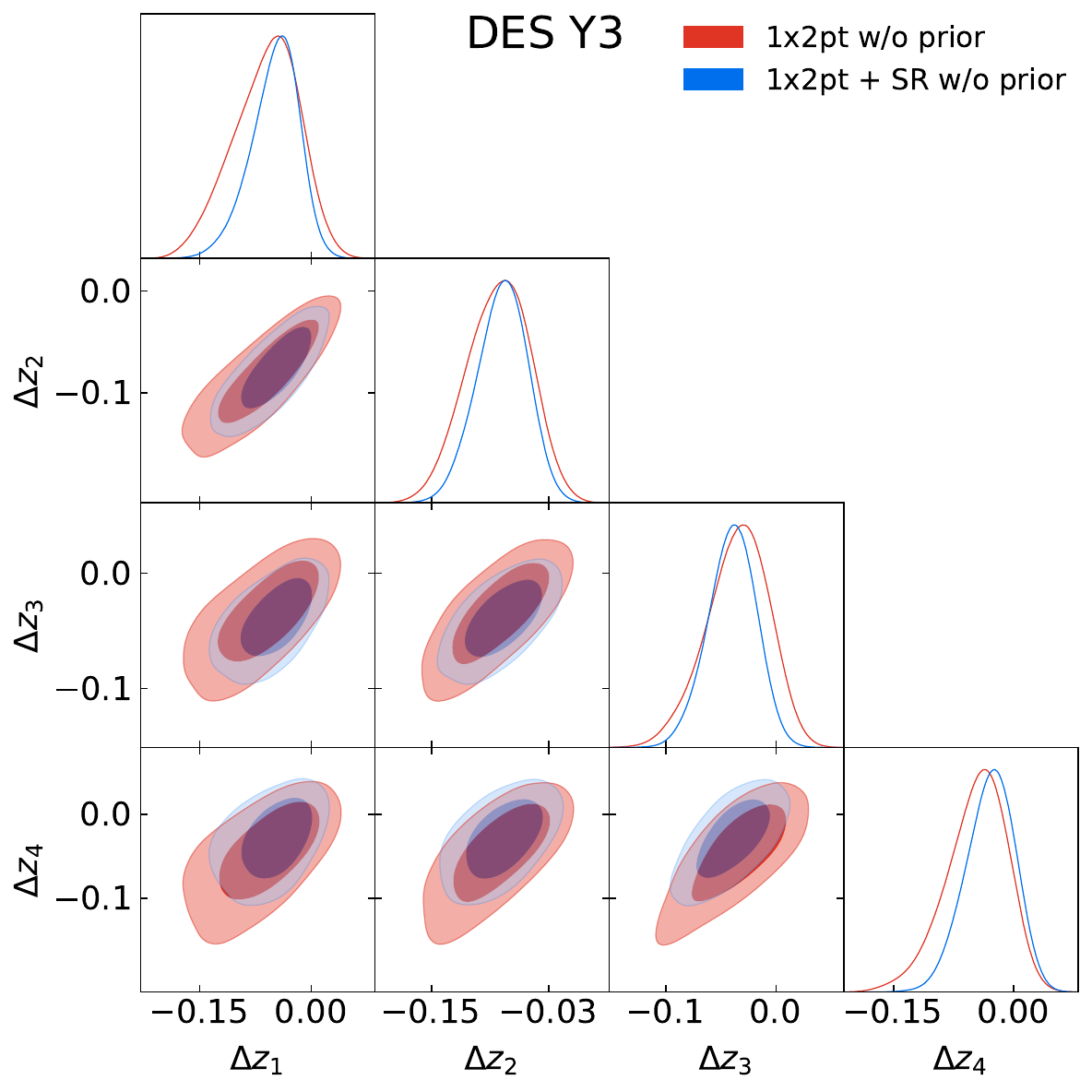}
    \vskip\baselineskip
    \includegraphics[width=0.9\linewidth]{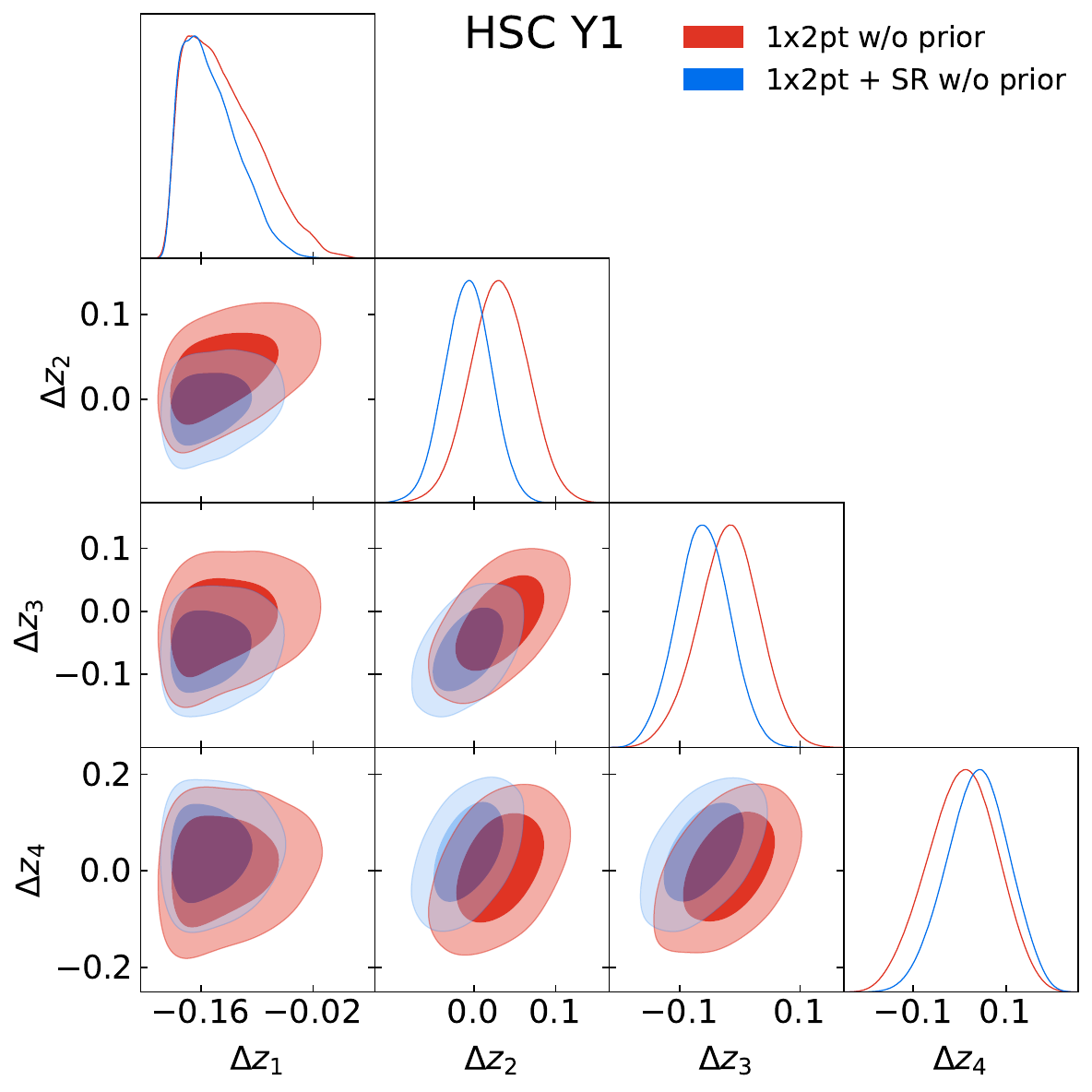}
    \caption{The joint confidence regions of $\Delta z$ parameters from MCMC fits varying cosmology and without the $\Delta z$ prior. }
    \label{MCMC_Dz}
\end{figure}

\begin{figure}
    \centering
    \includegraphics[width=0.9\linewidth]{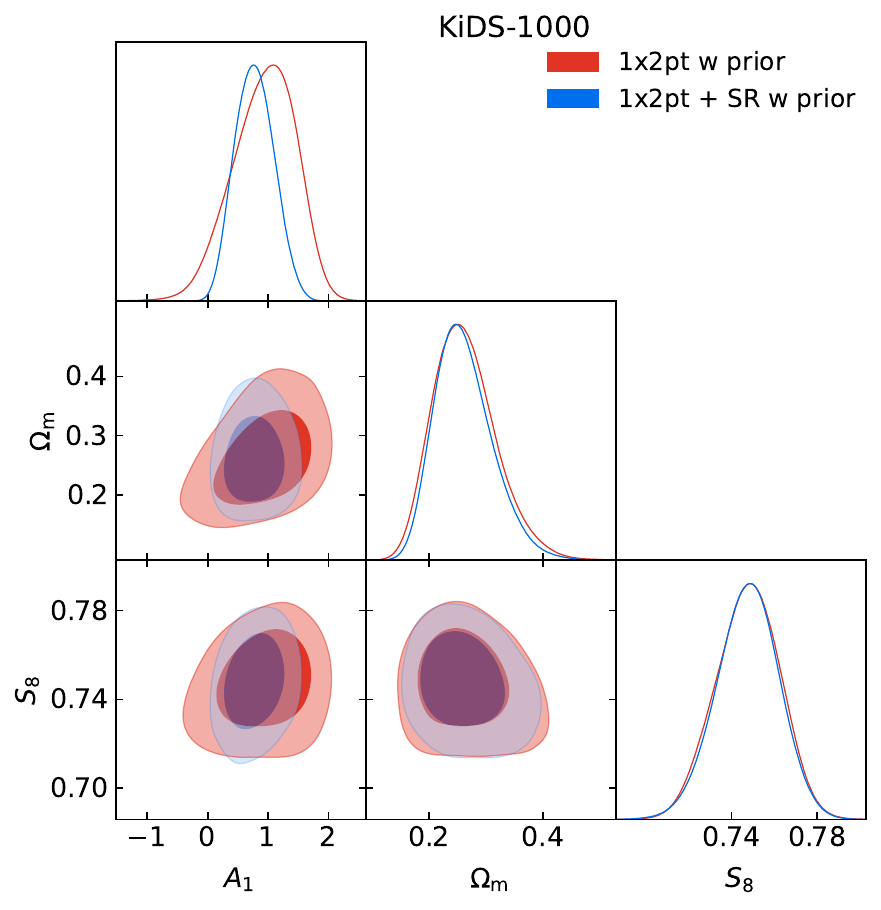}
    \vskip\baselineskip
    \includegraphics[width=0.87\linewidth]{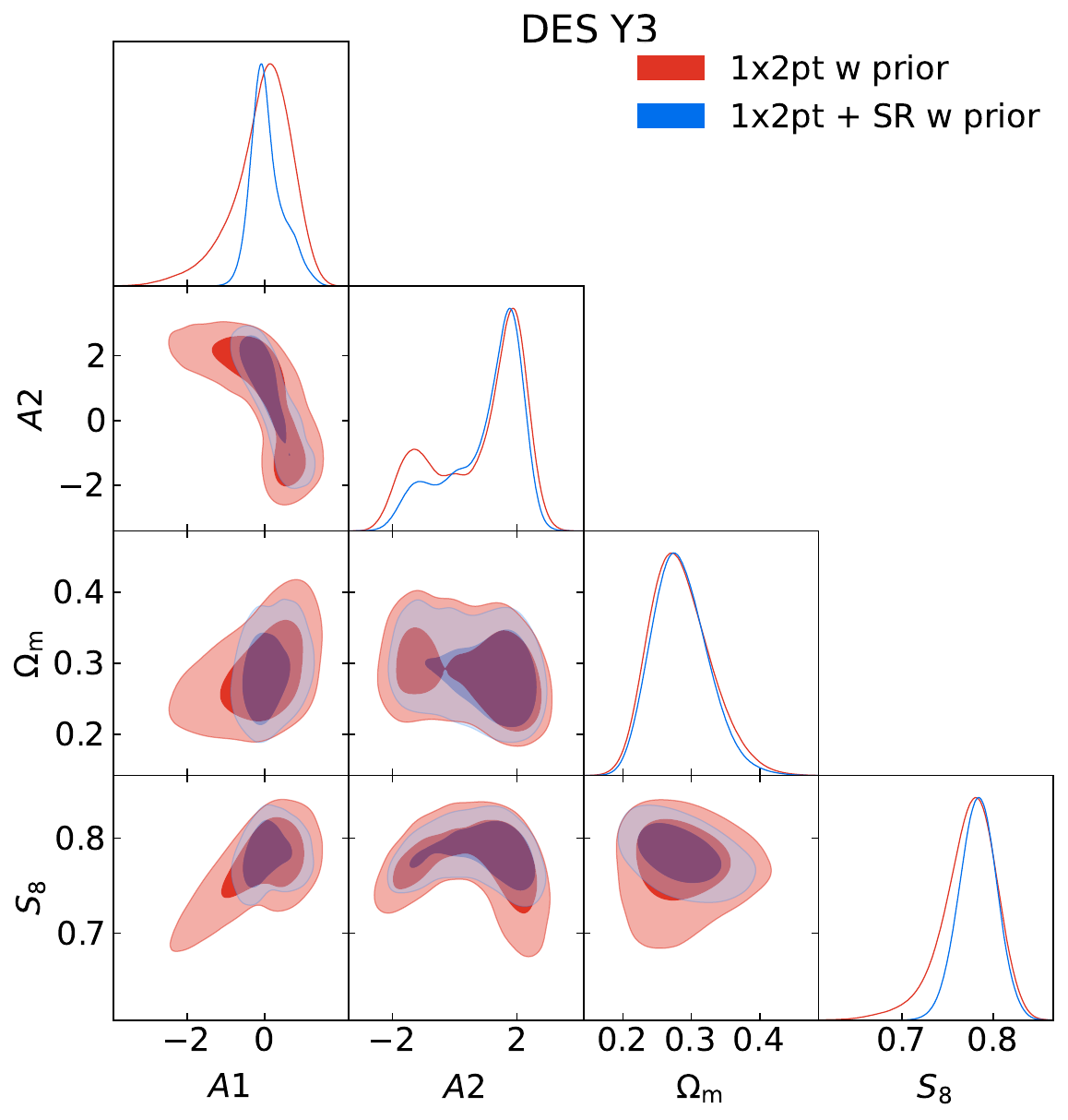}
    \vskip\baselineskip
    \includegraphics[width=0.87\linewidth]{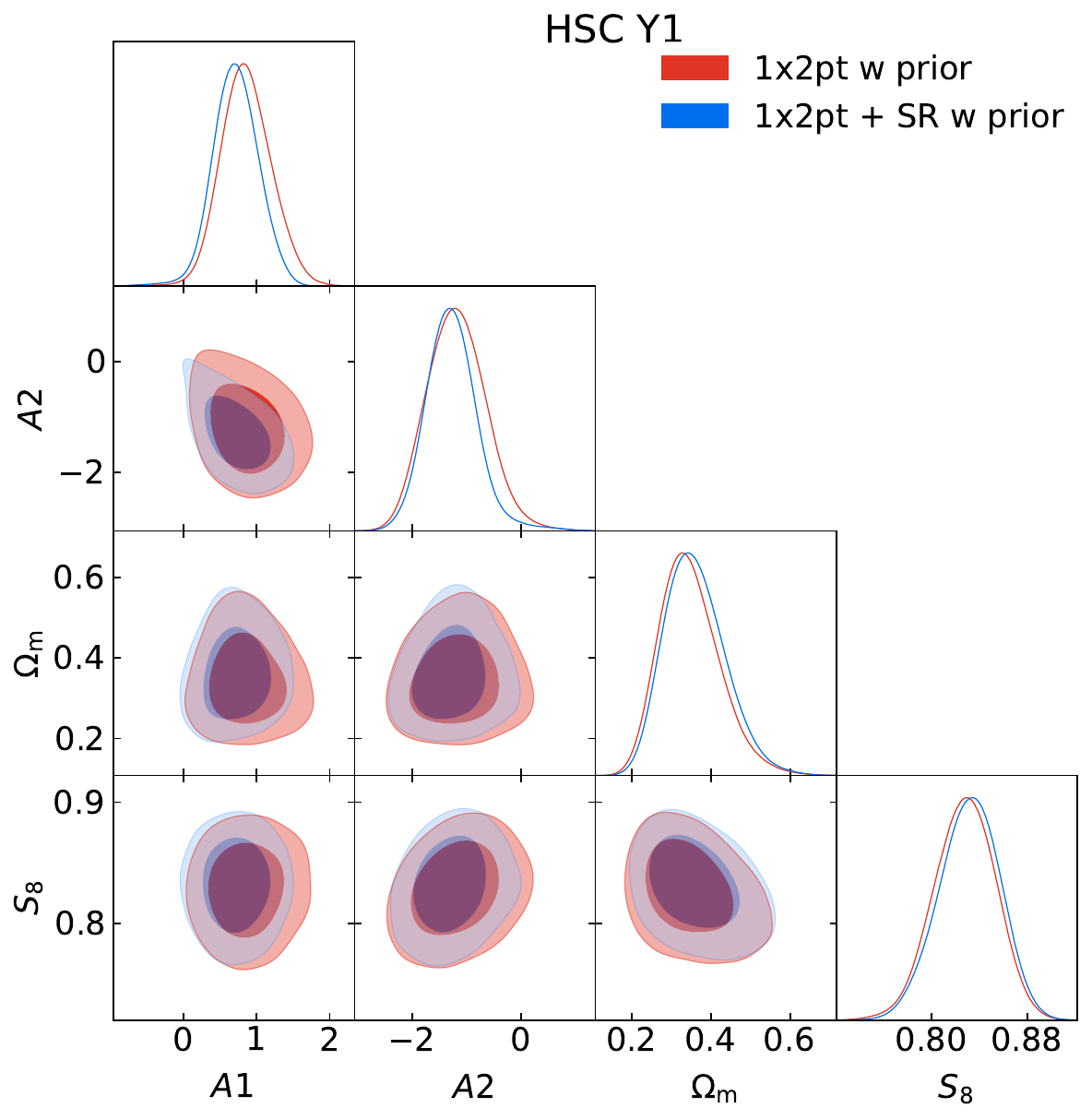}
    \caption{The joint confidence regions of IA and cosmological parameters from MCMC fits with the $\Delta z$ prior. }
    \label{MCMC_IA}
\end{figure}

When cosmological and lensing systematic parameters are simultaneously varied, we find more reliable improvements from adding shear ratio information to cosmic shear.  We illustrate the joint confidence intervals of the $\Delta z$ parameters in Figure \ref{MCMC_Dz}.  In some cases (most notably HSC), the inclusion of shear ratio data significantly improves the $\Delta z$ measurements if the prior is not available.  We note that for some tomographic bins in KiDS, the $\Delta z$ posterior encounters the edge of the wide uniform prior, effectively masking some of the improvement resulting from the addition of SR information.  The determination of the IA parameters consistently improves when adding SR data, with and without the $\Delta z$ prior.  The joint confidence intervals focussing on this parameter space are presented in Figure \ref{MCMC_IA}.  Turning to the cosmological parameters themselves: Table \ref{table:MC_result_improve} indicates that without the $\Delta z$ prior, the $S_8$ errors consistently improve by $5-15\%$; with this prior, we obtain significant improvements of 34\% for DES-Y3, although less so for KiDS-1000 and HSC-Y1.  In the DES-Y3 fit, there is more improvement because of the degeneracy between $S_8$ and the IA parameters in the TATT model.

\begin{figure}
  \includegraphics[width=\linewidth]{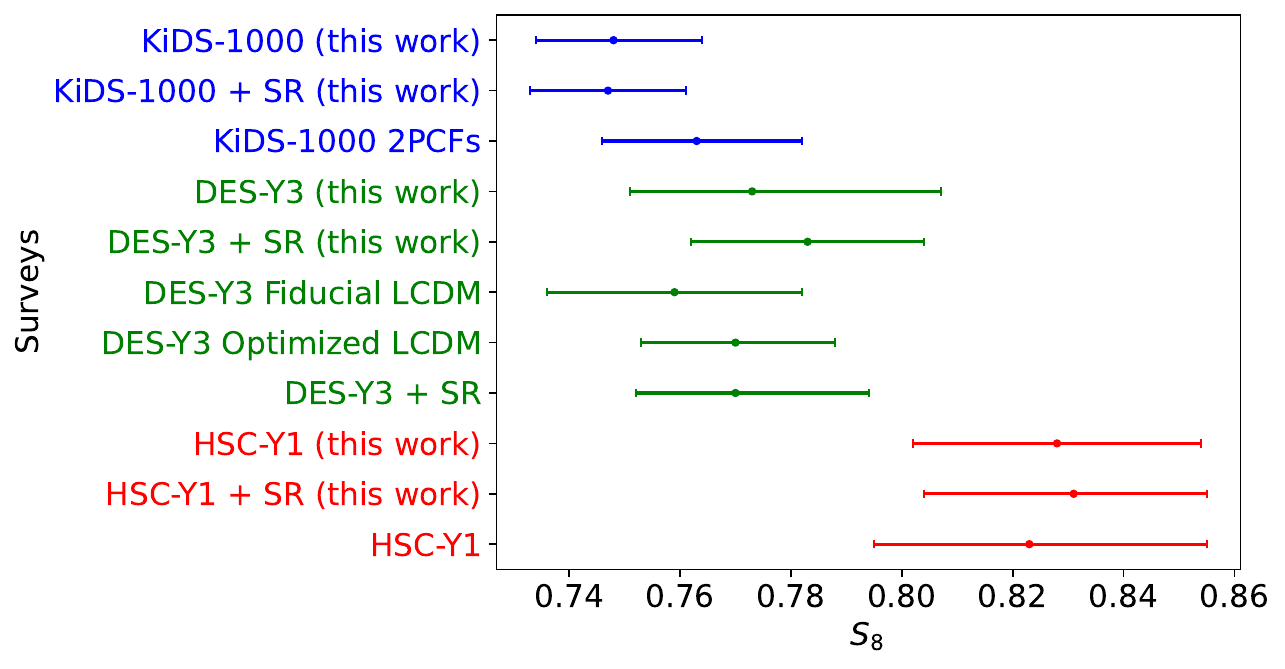}
  \caption{Comparison of $S_8$ measurements from our analysis of different datasets, and previous results presented by weak lensing survey collaborations. }
  \label{fig:compare_S8}
\end{figure}

As a cross-check of our analysis pipeline, we compared our cosmological parameter fits with the previous analyses of each weak lensing survey presented by \cite{kids1000_cut2}, \cite*{DES_cut}, \cite{2022PhRvD.105b3514A}, and \cite{hsc_cut} for KiDS-1000, DES-Y3 and HSC-Y1, respectively.  As displayed in Figure \ref{fig:compare_S8}, the results of our cosmic-shear-only fits agree well with previous analyses, with slight differences resulting from minor variations in the details of the modelling and analysis methods, none of which affect the conclusions of our work.  In particular there are small differences with respect to previous work in our modelling of the power spectrum, priors adopted when creating a unified analysis across the lensing surveys, and conventions for summarising the posterior probability distributions.  These differences are not important when assessing the potential benefits of adding SR to the analysis.



Comparing our results to the shear ratio analysis of DES-Y3 presented by \citet*{2022PhRvD.105h3529S}, we find similar resulting errors in the parameters but a somewhat different improvement after adding SR information.  We find an improvement of 8.6\% for $\Omega_\mathrm{m}$ and 33.9\% for $S_8$, whereas \citet*{2022PhRvD.105h3529S} obtained 3\% for $\Omega_\mathrm{m}$ and 25\% for $S_8$.  The likely principal source of these differences is that the galaxy-galaxy lensing measurements in \citet*{2022PhRvD.105h3529S} were based on photometric lenses identified in DES imaging, differing from our shear ratio measurements based on spectroscopic BOSS lenses.  Small modelling differences, such as in baryon feedback, may also affect the comparison. 

Summarising: the confidence regions in the multi-dimensional parameter space are complex, and the interactions do not always result in narrower posteriors.  The clearest cases are when all parameters are varying, in which the addition of shear ratio data consistently improves the determination of the IA amplitudes.  Generally, this results in a noticeable improvement in the $S_8$ value as well.

Finally, we review how these conclusions are affected when combining the shear ratio data with the full $3 \times 2$-pt correlations, rather than just with cosmic shear.  Similarly to Section \ref{sec:fisher}, we just consider the case including the $\Delta z$ prior and varying cosmology.  As we found in our Fisher matrix forecast, the improvements enabled by the shear ratio are much lower when the full $3 \times 2$-pt correlations are available.  For example, $\Delta z$ parameters and $S_8$ improve by a few percent, whilst IA amplitudes improve by up to $20\%$ for DES-Y3 but less significantly for KiDS and HSC. 


%

\begin{table*}
\centering
\begin{tabular} { c c c c c}
 Parameter &  Without $\Delta z$ prior & $+$SR & With $\Delta z$ prior & $+$SR \\
\hline
Fixed Cosmology \\
\hline
KiDS-1000\\
{$\Delta z_1     $} & $0.0292^{+0.023}_{-0.0067} $ &   $0.009^{+0.033}_{-0.021}   $ &  $0.0052\pm 0.0091          $ & $0.0011\pm 0.0091          $  \\

{$\Delta z_2     $} & $0.0478^{+0.014}_{-0.0070} $ &   $0.032^{+0.018}_{-0.013}   $ &  $0.0141\pm 0.0084          $ & $0.0094\pm 0.0083          $ \\

{$\Delta z_3     $} & $-0.0259^{+0.0077}_{-0.012}$ &  $-0.0363^{+0.0075}_{-0.0097}$ & $-0.0201^{+0.0087}_{-0.0077}$ & $-0.0266\pm 0.0082         $ \\

{$\Delta z_4     $} & $-0.0389^{+0.0027}_{-0.0071}$ & $-0.0380^{+0.0030}_{-0.0065}$ &  $-0.0242\pm 0.0063         $ & $-0.0245^{+0.0065}_{-0.0059}$ \\

{$\Delta z_5     $} & $-0.031^{+0.014}_{-0.023}  $ &  $-0.023^{+0.019}_{-0.022}  $  &  $-0.0045\pm 0.0076         $ & $-0.0031\pm 0.0080         $ \\
{$A_1$}             & $1.78^{+0.35}_{-0.30}      $ &  $1.10\pm 0.27              $  &  $1.45\pm 0.25              $ & $1.11\pm 0.22              $ \\

\hline

DES-Y3\\

{$\Delta z_1     $} & $-0.061^{+0.051}_{-0.038}  $ & $-0.049^{+0.037}_{-0.026}  $ & $0.005\pm 0.012             $ & $0.004\pm 0.012            $\\

{$\Delta z_2     $} & $-0.081^{+0.035}_{-0.031}  $ & $-0.078^{+0.028}_{-0.024}  $ & $-0.0205^{+0.0095}_{-0.0080}$ & $-0.0229^{+0.0095}_{-0.0082}$\\

{$\Delta z_3     $} & $-0.035^{+0.032}_{-0.025}  $ & $-0.040^{+0.023}_{-0.020}  $ & $0.0013\pm 0.0081           $ & $-0.0043\pm 0.0081         $\\

{$\Delta z_4     $} & $-0.047^{+0.044}_{-0.031}  $ & $-0.029^{+0.034}_{-0.028}  $ & $-0.001\pm 0.012            $ & $0.004\pm 0.012            $\\

{$A_1$}             & $-0.47^{+0.75}_{-0.55}     $  & $-0.39^{+0.49}_{-0.40}     $ & $0.44^{+0.25}_{-0.46}      $ & $0.26^{+0.18}_{-0.39}      $\\
{$A_2$}             &  $0.84^{+1.3}_{-0.65}       $ & $1.02^{+1.1}_{-0.51}       $ & $0.54^{+1.2}_{-0.72}       $ & $0.68^{+0.95}_{-0.61}      $\\
\hline

HSC-Y1\\
{$\Delta z_1     $} & $-0.132^{+0.026}_{-0.061}  $ & $-0.148^{+0.020}_{-0.046}  $ & $-0.047\pm 0.025           $ & $-0.060\pm 0.024           $ \\

{$\Delta z_2     $} & $0.031\pm 0.035            $ & $-0.008\pm 0.028           $ & $0.0082^{+0.0092}_{-0.010} $ & $0.002\pm 0.010            $ \\

{$\Delta z_3     $} & $-0.019\pm 0.050           $ & $-0.061\pm 0.044           $ & $-0.014\pm 0.024           $ & $-0.025\pm 0.023           $ \\

{$\Delta z_4     $} & $0.006\pm 0.074            $ & $0.039\pm 0.066            $ & $0.001\pm 0.029            $ & $0.015\pm 0.027            $ \\

{$A_1$}             & $0.21^{+0.67}_{-0.49}      $  & $0.11^{+0.41}_{-0.35}      $ & $0.75^{+0.32}_{-0.36}      $ & $0.52\pm 0.33              $ \\
{$A_2$}             &  $-0.99^{+0.59}_{-0.75}     $ & $-0.999^{+0.55}_{-0.91}    $ & $-1.31^{+0.52}_{-0.69}     $ & $-1.41^{+0.43}_{-0.75}     $ \\
\hline
\hline

Varying Cosmology\\
\hline
KiDS-1000\\
{$\Delta z_1     $} & $0.0241^{+0.027}_{-0.0092} $ & $0.007^{+0.033}_{-0.022}   $ & $0.0034\pm 0.0093          $ & $0.0009\pm 0.0091          $\\

{$\Delta z_2     $} & $0.0465^{+0.015}_{-0.0071} $ & $0.032^{+0.020}_{-0.013}   $ & $0.0099\pm 0.0094          $ & $0.0069\pm 0.0091          $\\

{$\Delta z_3     $} & $-0.017^{+0.011}_{-0.018}  $ & $-0.0315^{+0.0094}_{-0.013}$ & $-0.0152\pm 0.0092         $ & $-0.0214\pm 0.0089         $\\

{$\Delta z_4     $} & $-0.0232^{+0.0084}_{-0.021}$ & $-0.0248^{+0.0072}_{-0.018}$ & $-0.0145\pm 0.0075         $ & $-0.0148\pm 0.0075         $\\

{$\Delta z_5     $} & $0.004^{+0.027}_{-0.019}   $ & $0.0098^{+0.022}_{-0.014}  $ & $0.0069\pm 0.0086          $ & $0.0072\pm 0.0085          $\\
{$A_1$}             & $1.38^{+0.57}_{-0.41}      $ & $0.78^{+0.30}_{-0.35}      $ & $0.93^{+0.59}_{-0.44}      $ & $0.77^{+0.31}_{-0.34}      $\\
{$\Omega_\mathrm{m}       $} & $0.268^{+0.046}_{-0.064}   $ & $0.265^{+0.041}_{-0.062}   $ & $0.263^{+0.045}_{-0.062}   $ & $0.262^{+0.040}_{-0.058}   $\\
{$S_8$}             & $0.754\pm 0.019            $ & $0.751^{+0.018}_{-0.016}   $ & $0.748^{+0.016}_{-0.014}   $ & $0.747\pm 0.014            $\\
\hline

DES-Y3 \\
{$\Delta z_1     $} & $-0.038^{+0.052}_{-0.038}  $ & $-0.050^{+0.043}_{-0.037}  $ & $-0.003\pm 0.014           $ & $0.001\pm 0.012            $\\

{$\Delta z_2     $} & $-0.030^{+0.053}_{-0.038}  $ & $-0.075\pm 0.038           $ & $-0.0175^{+0.011}_{-0.0093}$ & $-0.0196^{+0.011}_{-0.0092}$\\

{$\Delta z_3     $} & $0.069^{+0.083}_{-0.047}   $ & $-0.021^{+0.053}_{-0.059}  $ & $0.0053\pm 0.0089          $ & $0.0000\pm 0.0088          $\\

{$\Delta z_4     $} & $0.082^{+0.10}_{-0.043}    $ & $0.004\pm 0.078            $ & $0.004\pm 0.013            $ & $0.009\pm 0.013            $\\

{$A_1$}             & $-0.73^{+0.93}_{-0.70}     $ & $-0.28\pm 0.44             $ & $-0.07^{+0.89}_{-0.50}     $ & $0.06^{+0.28}_{-0.51}      $\\
{$A_2$}             & $-0.2\pm 1.6               $ & $0.64^{+1.4}_{-0.79}       $ & $0.8^{+1.8}_{-2.4}         $ & $1.0^{+1.5}_{-0.52}        $\\

{$\Omega_\mathrm{m}       $} & $0.304^{+0.045}_{-0.067}   $ & $0.338^{+0.047}_{-0.067}   $ & $0.285^{+0.036}_{-0.051}   $ & $0.284^{+0.035}_{-0.045}   $\\
{$S_8$}             & $0.700^{+0.039}_{-0.065}   $ & $0.781\pm 0.049            $ & $0.773^{+0.034}_{-0.022}   $ & $0.783\pm 0.021            $\\

\hline
HSC-Y1 \\
{$\Delta z_1     $} & $-0.137^{+0.022}_{-0.057}  $ & $-0.153^{+0.018}_{-0.042}  $ & $-0.047^{+0.027}_{-0.022}  $ & $-0.058^{+0.024}_{-0.021}  $\\

{$\Delta z_2     $} & $0.024\pm 0.062            $ & $-0.049^{+0.033}_{-0.039}  $ & $0.005\pm 0.010            $ & $-0.001\pm 0.010           $\\

{$\Delta z_3     $} & $-0.015\pm 0.084           $ & $-0.113^{+0.036}_{-0.065}  $ & $-0.016\pm 0.025           $ & $-0.028\pm 0.024           $\\

{$\Delta z_4     $} & $0.017^{+0.13}_{-0.089}    $ & $-0.031^{+0.073}_{-0.11}   $ & $-0.002\pm 0.029           $ & $0.014\pm 0.028            $\\
{$A_1$}             & $0.38^{+0.64}_{-0.41}      $ & $0.36\pm 0.38              $ & $0.86^{+0.32}_{-0.36}      $ & $0.71\pm 0.32              $\\
{$A_2$}             & $-1.03\pm 0.61             $ & $-0.90^{+0.40}_{-0.55}     $ & $-1.19\pm 0.55             $ & $-1.27^{+0.40}_{-0.48}     $\\

{$\Omega_\mathrm{m}       $} & $0.361^{+0.076}_{-0.11}    $ & $0.413^{+0.084}_{-0.13}    $ &  $0.347^{+0.061}_{-0.086}   $ & $0.360^{+0.064}_{-0.086}   $\\
{$S_8$}             & $0.811\pm 0.049            $ & $0.856^{+0.046}_{-0.038}   $  & $0.828\pm 0.026            $ & $0.831^{+0.027}_{-0.024}   $\\
\hline
\end{tabular}
\caption{Full results of our MCMC fits to the cosmic shear correlation functions of KiDS-1000, DES-Y3 and HSC-Y1, including and excluding the shear ratio dataset, and the prior for $\Delta z$.  We include results from fixed cosmology and varying cosmology fits. We quote the mean of the posterior and the upper and lower intervals comprising the 68\% credible region.} 
\label{error_mc}
\end{table*}

\begin{table*}
\centering
\begin{tabular} { c c c c c c c}
\hline
 Case & $\chi^2$ & $\chi^2$ & $\chi^2_\mathrm{red}$ & $\chi^2_\mathrm{red}$ & $N_\mathrm{data}$ & $N_\mathrm{par}$\\
& Fix Cos. & Vary Cos. & Fix Cos. & Vary Cos. & \\
\hline
\textbf{KiDS-1000} \\
without $\Delta z$ prior      &   257.95 &  251.81 & 1.15 & 1.12 & 225 & 12\\
without $\Delta z$ prior + SR &   292.17 &  281.68 & 1.19 & 1.15 & 245 & 12\\
with $\Delta z$ prior              &   268.96 &  258.78 & 1.20  & 1.15 & 225 & 18\\ 
with $\Delta z$ prior + SR         &   299.26 &  286.99 & 1.22 & 1.17 & 245 & 18\\
\textbf{DES-Y3} \\
without $\Delta z$ prior      &   237.70 &   238.70 & 1.05 & 1.05 & 227 & 13\\
without $\Delta z$ prior + SR &   252.84 &   253.66 & 1.04 & 1.05 & 242 & 13\\
with $\Delta z$ prior              &   239.91 &   242.01 & 1.06 & 1.07 & 227 & 19\\
with $\Delta z$ prior + SR         &   257.81 &   257.12 & 1.07 & 1.06 & 242 & 19\\
\textbf{HSC-Y1} \\
without $\Delta z$ prior      &   212.12 &  209.93 & 1.25 & 1.23 & 170 & 13\\
without $\Delta z$ prior + SR &   231.65 &  225.86 & 1.25 & 1.22 & 185 & 13\\
with $\Delta z$ prior              &   213.46 &  210.36 & 1.26 & 1.24 & 170 & 19\\
with $\Delta z$ prior + SR         &   233.99 &  229.35 & 1.26 & 1.24 & 185 & 19\\

\hline
\end{tabular}
\caption{The $\chi^2$ values for our fits to cosmic shear and shear ratio datasets, for fixed cosmology and varying cosmology.  For convenience, we also display the corresponding values of the reduced $\chi^2$ statistic ($\chi^2_\mathrm{red}$) assuming that the number of degrees of freedom is given by the difference between the number of data points ($N_\mathrm{data}$) and the number of fitted parameters ($N_\mathrm{par}$).}
\label{chi_mcmc}
\end{table*}

\begin{table*}
\centering
\begin{tabular}{ccccc}
\hline
WL survey & Parameter & Cosmology & Improvement & Improvement \\
& & & without $\Delta z$ prior & with $\Delta z$ prior  \\
\hline

KiDS-1000 & $\Delta z$ & Fixed & -37.1 - 11.6\% & -4.7 - 3.0\% \\
DES-Y3    & $\Delta z$ & Fixed & 20.1 - 27.1\%  & 0.9 - 3.3\% \\
HSC-Y1    & $\Delta z$ & Fixed & 10.1 - 24.1\%  & -1.3 - 8.0\% \\
                
KiDS-1000 & $A_1$ & Fixed & 16.9\% & 10.7\% \\
DES-Y3    & $A_1,A_2$ & Fixed & 31.1, 8.9\%    & 12.3, 13.4\% \\
HSC-Y1    & $A_1,A_2$ & Fixed & 34.6, -13.3\%  & 8.2, -9.3\% \\

\hline

KiDS-1000 & $\Delta z$ & Varying & -35.4 - 20.7\% & 0.5 - 3.6\% \\
DES-Y3    & $\Delta z$ & Varying & 1.3 - 18.0\%  & -0.2 - 8.5\% \\
HSC-Y1    & $\Delta z$ & Varying & 9.0 - 41.6\%  & 2.58 - 7.0\% \\
                
KiDS-1000 & $A_1$ & Varying & 36.2\%  & 38.8\% \\
DES-Y3    & $A_1,A_2$ & Varying & 45.9, 32.9\% & 45.4, 16.7\% \\
HSC-Y1    & $A_1,A_2$ & Varying & 38.1, 14.5\% & 9.21, 10.94\% \\

\hline

KiDS-1000 & $\Omega_\mathrm{m}$ & Varying & 4.0\%   & 7.8\% \\
DES-Y3    & $\Omega_\mathrm{m}$ & Varying & 0.4\%    &  8.6\% \\
HSC-Y1    & $\Omega_\mathrm{m}$ & Varying & -15.2\%  & -0.1\% \\
                
KiDS-1000 & $S_8$ & Varying & 5.1\% & 2.4\% \\
DES-Y3    & $S_8$ & Varying & 13.2\% & 33.9\% \\
HSC-Y1    & $S_8$ & Varying & 15.1\% & 1.3\%  \\

\hline
\end{tabular}
\caption{The improvement in the determination of selected model parameters from MCMC fitting, following the addition of shear ratio (SR) data to cosmic shear ($1 \times 2$-pt) data.  Inclusion of the prior in $\Delta z$ (from photometric redshift calibration) is also significant in determining the relative contribution of shear ratio data.  We indicate if the cosmological parameters are held fixed or varying in each analysis.}
\label{table:MC_result_improve}
\end{table*}

\section{Discussion and Conclusions}
\label{sec:conc}

In this paper we have considered the utility of the shear ratio test to improve determinations of cosmological and astrophysical parameters, when fit to weak lensing and large-scale structure datasets.  For the analysis, we used lenses from the Baryon Oscillation Spectroscopic Survey that overlap with three important weak lensing surveys: KiDS-1000, DES-Y3 and HSC-Y1. These overlapping datasets allow us to perform analyses of two-point correlation statistics: $1 \times 2$-pt (cosmic shear) and $3 \times 2$-pt (cosmic shear, galaxy-galaxy lensing, and galaxy clustering).

The shear ratio test provides potential additional information to these correlation functions, originating from the relative amplitude of galaxy-galaxy lensing signals [$\gamma_\mathrm{t}(\theta)$] from different source populations around the same lenses.  This ratio potentially reduces the sensitivity to non-linear effects and allows smaller scales to be accessed.  In our work we addressed the following questions:

\begin{enumerate}

    \item What constraints do shear ratios provide on the mean redshifts of the source galaxies (using the offset $\Delta z$ parameters), and how do these compare with estimates of $\Delta z$ provided by photometric redshift calibration techniques?
    
    \item What constraints do shear ratios provide on the intrinsic alignment parameters?
    
    \item Can shear ratios improve constraints on the $S_8$ parameter?
    
\end{enumerate}

To answer these questions we considered both a comprehensive Fisher matrix analysis, together with a Bayesian inference analysis using MCMC, based on these datasets.  We apply a new Bayesian fitting method to measure the shear ratio \citep[inspired by][]{bayes_fit}, avoiding biases which can be introduced when the denominator of a ratio approaches zero.  The availability of three source datasets allows us to inter-compare results for different weak lensing surveys, using the same lens galaxies.

We summarise the conclusions of our aims as follows:

\begin{enumerate}

    \item The addition of shear ratio information to cosmic shear allows the mean redshifts of the source samples to be determined significantly more accurately.  The additional constraining power represented by the shear ratio data is somewhat weaker than the prior in $\Delta z$ produced by photometric redshift calibration techniques, but can nonetheless provide a useful cross-check of these priors.
    
    \item Inclusion of shear ratio data consistently improves the determination of the IA parameters, when both cosmological and astrophysical parameters are jointly varied in the analysis.  These improvements are significant when SR data is combined with cosmic shear data only (even when a $\Delta z$ prior is available), although less significant when the full $3 \times 2$-pt correlations are available.

    \item Through improving the constraints on the astrophysical systematic parameters, inclusion of shear ratio data consistently benefits the determination of $S_8$, when combined with cosmic shear data only.  However, these improvements are less significant when shear ratio data is combined with the full $3 \times 2$-pt correlations.
    
\end{enumerate}

In general, our results agree well with the previous shear ratio analysis for DES-Y3 presented by \citet*{2022PhRvD.105h3529S}.  An important difference to highlight is that \citet*{2022PhRvD.105h3529S} generated shear ratio measurements using photometric lenses selected across the whole DES-Y3 footprint, whereas our analysis uses spectroscopic lenses within the smaller area overlapped by BOSS.  In future, shear ratio analyses will be significantly enhanced by upcoming data from the Dark Energy Spectroscopic Instrument (DESI), which will increase the overlap of galaxy redshift and lensing surveys, enabling better cross-checks of lensing systematic parameters and determinations of $S_8$.  Weak lensing datasets will also be extended in the future through projects such as the {\it Euclid} satellite, the Rubin Observatory's Legacy Survey of Space and Time (LSST), and the Roman Space Telescope.  We can also consider adding further external datasets, such as CMB lensing.

\section*{Acknowledgements}

We thank the anonymous referee for thorough comments on the original manuscript.  NE would like to acknowledge the financial support received through a Swinburne University Postgraduate Research Award. AP acknowledges support from the European Union’s Horizon Europe program under the Marie Skłodowska-Curie grant agreement 101068581. We are grateful to Agne Semenaite for useful comments on this work.

\section*{Data availability}

The data underlying this article will be shared on reasonable request to the corresponding author.

\bibliographystyle{mnras}
\bibliography{template}

\end{document}